\documentclass[%
reprint,
 superscriptaddress,
 amsmath,amssymb,
 aps,
pra,
floatfix,
longbibliography]{revtex4-1}

\usepackage{graphicx}% Include figure files
\usepackage{subfigure}
\usepackage{dcolumn}% Align table columns on decimal point
\usepackage{bm}% bold math
\usepackage{epstopdf}
\usepackage{xcolor}
\usepackage{hyperref}% add hypertext capabilities
\usepackage{appendix} % ADDED 6/13/2019
\usepackage{mathtools}
\usepackage{tablefootnote}
\usepackage{comment}
\usepackage{textcase}
\usepackage{soul}
\usepackage{float}
\usepackage[normalem]{ulem}
\usepackage[english]{babel}
\begin{document}
\setstcolor{red}
%\preprint{APS/123-QED}

\title{Three and four identical fermions near the unitary limit}% Force line breaks with \\

\author{Michael D. Higgins}
\affiliation{Department of Physics and Astronomy, Purdue University, West Lafayette, Indiana 47907 USA}
\author{Chris H. Greene}
\affiliation{Department of Physics and Astronomy, Purdue University, West Lafayette, Indiana 47907 USA}
\affiliation{Purdue Quantum Science and Engineering Institute, Purdue University, West Lafayette, Indiana 47907 USA}

\date{\today}% It is always \today, today,

%\begin{abstract}
%Low-energy elastic and inelastic scattering in the Ps(1$s$)-Ps(2$s$) channel is treated in a four-body hyperspherical coordinate calculation. Adiabatic potentials are calculated for triplet-triplet, singlet-singlet, and singlet-triplet spin symmetries in the spin representation of coupled electrons and coupled positrons, with total angular momentum $L=0$ and parity equal to $+1$. The s-wave scattering lengths for the asymptotic Ps(1$s$)-Ps(2$s$) channel are calculated for each spin configuration. Results obtained for the s-wave scattering lengths are $a_{\mathrm{TT}}=$~$7.3(2)a_0-i0.02(1)a_0$, $a_{\mathrm{SS}}=$~$13.2(2)a_0-i0.9(2)a_0$, and $a_{\mathrm{ST}}=$~$9.7(2)a_0$ for each spin configuration. Spin recoupling is implemented to extract the scattering lengths for collisions of Ps in different spin configurations through properly symmetrized unitary transformations. Calculations of experimentally relevant scattering lengths and cross-sections are carried-out for Ps atoms initially prepared in different uncoupled spin states.
%\end{abstract}

\begin{abstract}
    This work analyzes the three and four equal--mass fermionic systems near and at the $s$-- and $p$--wave unitary limits using hyperspherical methods. The unitary regime addressed here is where the two--body dimer energy is at zero energy. For fermionic systems near the $s$--wave unitary limit, the hyperradial potentials in the $N$--body continuum exhibit a universal long--range $R^{-3}$ behavior governed by the $s$--wave scattering length alone. The implications of this behavior on the low energy phase shift are discussed. At the $p$--wave unitary limit, the four--body system is studied through a qualitative look at the structure of the hyperradial potentials at unitarity for the $L^{\pi}=0^{+}$ symmetry. A quantitative analysis shows that there are tetramer states in the lowest hyperradial potentials for these systems. Correlations are made between these tetramers and the corresponding trimers in the two--body fragmentation channels. Universal properties related to the four--body recombination process $\mathrm{A+A+A+A}\leftrightarrow \mathrm{A_3+A}$ are discussed.
\end{abstract}

\pacs{Valid PACS appear here}% PACS, the Physics and Astronomy
\maketitle

\section{Introduction}
Few--body physics has been extensively studied in a regime where the two--body $s$--wave interaction is at the unitary limit, i.e. when $a_s\rightarrow\infty$. One subset of systems of interest are interacting bosons, which have been studied as a way to probe universality and to understand the manifestations of the three--body Efimov effect \cite{osti_4068792,EFIMOV1973157,NaidonReview}. Fermionic systems have also been studied extensively in recent years. Some of these studies include interactions of two--component Fermi gases near unitarity, in connection with the Bose--Einstein Condensate to Bardeen–-Cooper–-Schrieffer (BEC--BCS) crossover problem \cite{regal2007,PhysRevLett.99.090402,vonStecher2007prl,PhysRevLett.97.150401,Rittenhouse_2011,PhysRevLett.94.070403}. Other works include recent experiments on fermionic atoms, such as $\prescript{40}{}{\mathrm{K}}$ and $\prescript{6}{}{\mathrm{Li}}$, near $p$--wave Feshbach resonances whose interactions are tuned via magnetic fields \cite{PhysRevLett.120.133401,PhysRevA.98.020702,PhysRevA.99.052704}.

In a recent study published in \cite{PhysRevA.105.013308}, systems consisting of three equal--mass fermions interacting through a Lennard--Jones potential was studied to investigate the role $p$--wave unitary interactions have on the three body system, and also to see whether there are signatures of an Efimov effect. As a result of this study, it has become clear that there is no Efimov effect for equal--mass fermionic systems in 3D,  shown using the Born--Oppenheimer framework. In the hyperspherical Born-Oppenheimer approximation, the equal--mass fermionic system has a lowest potential with a long--range attractive $1/R^2$ tail that resembles an Efimov--like potential. However, when the second--derivative diagonal adiabatic correction is included, the long--range attraction that appears to cause an Efimov effect is cancelled, which confirms the nonexistence of the Efimov effect in these systems \cite{dIncaoPrivateCommun}. Instead there is only one universal trimer state at the $p$--wave unitary limit for symmetries $L^{\pi}=1^{-}$ and $1^{+}$.  

One interest in pursuing this work for $s$--wave interactions near the unitary limit stems from two recent papers published in \cite{PhysRevLett.125.052501,PhysRevC.103.024004} that investigated low energy scattering of a few interacting neutrons in the context of few--body nuclear physics. The neutron--neutron scattering length is negative and large compared to the range of the interaction $a/r_0\sim-19$, thus making few neutron systems good candidate systems for the study of near--unitary physics. In reference \cite{PhysRevC.103.024004}, as a result of studying low--energy scattering in the $N$-body continuum for different nuclear interactions, it was found the long--range form of the Born--Oppenheimer potentials contained a $1/R^3$ term that is universal and only dependent on the scattering length and symmetry. As a result of the long--range form of the potential, this leads to a low--energy enhancement in the density of states that was used to suggest a possible explanation of the 2016 experimental result of Kisamori et al. \cite{Kisamori} These results provide motivation to further look at this type of long--range behavior near the unitary limit for other systems, such as ultracold atomic systems with tunable interactions.

Previous work was able to relate four--body processes and properties to the universal Efimov trimers in the four--boson spectrum \cite{StecherNature,PhysRevLett.108.073201,Zenesini_2013,PhysRevLett.122.143001,PhysRevA.90.012502,Platter2004}. In Ref. \cite{StecherNature}, the four equal--mass boson system interacting at the $s$--wave unitary limit was studied in the hyperspherical framework to predict the rate of the four--body recombination process $B+B+B+B\leftrightarrow B_3+B$, later connected to the Innsbruck experimental data on Cs 4-atom recombination in an ultracold quantum gas \cite{NatureKramer2006,Zenesini_2013}. In particular, the universal ratio of the scattering length $a_c^{(4)}$ where a tetramer becomes bound to the corresponding scattering length $a_c^{(3)}$ for the formation of an Efimov trimer allowed theory to predict a resonant enhancement in the recombination rate at that scattering length. Other studies addressed interacting bosonic systems at the unitary limit for $N>4$ \cite{Zenesini_2013,PhysRevLett.107.200402,von_Stecher_2010,PhysRevA.74.063604,PhysRevLett.113.213201,PhysRevA.90.012502} to explore universal relations between higher $N$--body bound states and to provide insights into mechanisms for higher $N$--body recombination loss rates.

This article is organized as follows. Section \ref{sec:section1} gives an overview of the theoretical methods used in this work, mainly highlighting features of the Born--Oppenheimer method, as well as giving details on the two--body interactions used. In Secs. \ref{sec:section2} and \ref{sec:section3}, the main results of this work are presented for both $s$--wave and $p$--wave interactions near and at the unitary limit. In Sec. \ref{sec:section2}, the three-- and four--body potential energy curves as functions of the hyperradius are analyzed near and at the $s$--wave unitary limit, where the long--range behavior of the lowest hyperradial potentials are characterized. The implications of the long--range behavior is discussed through an analysis of the elastic phase shift in the three--body and four--body continua. Sec. \ref{sec:section3} treats the interactions of four fermions near the $s$--wave and $p$--wave unitary limits through analysis of the hyperradial potentials for the $0^{+}$ symmetry, providing insights into the relation between the universal tetramer states and their correlations with the universal trimer states. Lastly, a universal ratio related to trimer and tetramer formation is characterized and shown to provide a useful parameter for future four--body recombination measurements.

\section{Theoretical Methods\label{sec:section1}}
\subsection{Born--Oppenheimer approach}

Here we give an overview of the hyperspherical framework used in this work, which was previously described in \cite{PhysRevC.103.024004}. Within the Born--Oppenheimer approach the three-- and four--body systems are solved using an explicitly correlated Gaussian (ECG) basis in conjunction with the hyperspherical framework (CGHS) \cite{Varga1998,SV3,Rittenhouse_2011,RevModPhys.85.693}. The Hamiltonian for the systems considered here allows for separation of the center of mass coordinates from the relative coordinates, i.e. $\hat{H}=\hat{H}_{\mathrm{CM}}+\hat{H}_{\mathrm{rel.}}$. The center of mass Hamiltonian, $\hat{H}_{\mathrm{CM}}$, consists only of the kinetic energy operator of the center of mass and will be ignored in the following. The hyperradial and hyperangular kinetic energy operators, along with the particle--particle interactions are treated in the Hamiltonian of the relative coordinates, $\hat{H}_{\mathrm{rel.}}$.

The hyperangular kinetic energy and interaction energy operators together construct the adiabatic Hamiltonian. In hyperspherical coordinates, the generalized $N$--body adiabatic eigenvalue problem needed to be solved is:

\begin{equation}
    \label{eq:supp_eigvalueprob}
    H_{ad}(R,\Omega)\Phi_\nu(R,\Omega)=U_\nu(R)\Phi_\nu(R,\Omega)
\end{equation}
where $R$ is the hyperradius, $\Omega$ is a set of hyperangles, $\nu$ is an index that labels the eigenstates $\Phi_{\nu}(R,\Omega)$ of $H_{\mathrm{ad}}(R,\Omega)$, $U_{\nu}(R)$ are eigenvalues of $H_{\mathrm{ad}}$ at fixed $R$ that represent Born--Oppenheimer potentials, and
\begin{multline}
    \label{eq:adexpression}
    H_{ad}(R,\Omega)=\frac{\hbar^2}{2\mu R^2}\biggr[\Lambda^2+\frac{(3N-4)(3N-6)}{4}\biggr]\\+V_{\mathrm{int.}}(R,\Omega)
\end{multline}
where the operator $\Lambda^2$ represents the squared hyperangular grand--angular momentum of the system and $V_{\mathrm{int.}}(R,\Omega)$ represents the sum of two--body potential operators between the particles. The parameter $\mu$ is the hyperradial reduced mass, defined as $\mu=~(m_{1}m_{2}\cdot\cdot\cdot m_{N}/(m_{1}+m_{2}+\cdot\cdot\cdot+m_{N}))^{{1/(N-1)}}$ \cite{Rittenhouse_2011}. The hyperradius is typically treated as an adiabatic parameter and is defined in general as a coordinate proportional to the square root of the trace of the moment--of--inertia tensor.  There is an arbitrariness in the overall scaling of the mass--weighted Jacobi vectors used to construct it, however. Typically in atomic systems, the reduced mass used in the hyperspherical framework is the hyperspherical reduced mass given in \cite{Rittenhouse_2011}. However, when applying the hyperspherical framework to few--body systems in other fields such as nuclear physics, the reduced mass of choice often used is the nucleon--nucleon two--body reduced mass.

The adiabatic Hamiltonian given by Eq. \eqref{eq:adexpression} is diagonalized using the CGHS basis. In the non--interacting case, the eigenstates $\Phi_\nu(R,\Omega)$ describe the hyperspherical harmonics (HH) with the label $\nu$ corresponding to the HH quantum number $K$ (\cite{Rittenhouse_2011} and references therein). Furthermore,  there is a one-to-one correspondence between the non--interacting eigenstates $\Phi_\nu(R,\Omega)$ and the states of a $d$--dimensional isotropic harmonic oscillator, which helps understand why the non--interacting eigenstates exhibit high degeneracies (for an example, see \cite{PhysRevB.92.125427}).

The $N$--body wavefunction is expanded in the relative coordinates in the eigenstates of Eq. \eqref{eq:supp_eigvalueprob}, given by the ansatz
\begin{equation}
    \label{eq:ansatz}
    \Psi_{E}(R,\Omega)=R^{-\frac{3N-4}{2}}\sum_\nu F_{E,\nu}(R)\Phi_{\nu}(R,\Omega).
\end{equation}
The factor $(3N-4)(3N-6)/4$ in Eq. \eqref{eq:adexpression} comes from the multiplying factor of $R$ in Eq. \eqref{eq:ansatz}, which eliminates the first derivative acting on $F_{E,\nu}(R)$ in the hyperradial kinetic energy. Computing the quantity $<\Phi_{\mathrm{\nu^{\prime}}}|\hat{H}_{\mathrm{rel.}}|\Psi_{E}>$, where the operation $<\cdot>$ indicates integrating over the hyperangular coordinates and tracing over spin degrees of freedom, leads to the following coupled hyperradial Schr\"odinger equations,

\begin{multline}
    \label{eq:coupled}
    \left(-\frac{\hbar^2}{2\mu}\frac{\partial^2}{\partial{R^2}}+W_{\nu}(R)-E\right)F_{\mathrm{E,\nu}}\left(R\right)\\
    -\frac{\hbar^2}{2\mu}\sum_{\nu^\prime\neq\nu}\left(2P_{\nu\nu^\prime}(R)\frac{\partial}{\partial{R}}+Q_{\nu\nu^\prime}(R)\right)F_{E,\nu^\prime}(R)=0
\end{multline}
where $P_{\nu \nu^{'}}(R)=\biggr<\Phi_{\mathrm{\nu^{\prime}}}\biggr|\frac{\partial\Phi_{\nu}}{\partial R}\biggr>$ and $Q_{\nu \nu^{'}}(R)=\biggr<\Phi_{\mathrm{\nu^{\prime}}}\biggr|\frac{\partial^2\Phi_{\nu}}{\partial R^2}\biggr>$ are first derivative and second derivative non--adiabatic couplings, and $W_{\nu}(R)=U_{\nu}(R)-\frac{\hbar^{2}}{2\mu}Q_{\nu\nu}(R)$ is the $\nu^{\mathrm{th}}$ effective adiabatic potential \cite{Wang, e_+e_-Daily,DailyAsym}.

\subsection{Two--body interactions}
The Hamiltonian considered in this work consists entirely of two--body interactions. Throughout, the values of $\hbar$ and $m$ are chosen to be in units where $\hbar=1$ and $m=1$. Two different types of interactions are used here:  either short--range interactions or else a long--range Van der Waals interaction with a short--range cutoff. In utilizing the explicitly--correlated Gaussian basis, the most convenient form of the interaction is that of a Gaussian interaction or a Gaussian multiplied by some arbitrary power. More explicitly, there are four different types of short--range two--body interactions considered in this work, presented in Eqs. \eqref{eq:interactions},
\begin{subequations}
\label{eq:interactions}
%\begin{equation}
%    \label{eq:interactions1}
%     \textcolor{red}{V(r)=\alpha_0\biggr(\frac{\hbar^2}{2\%mu_{\mathrm{2B}} r_0^2}\biggr)e^{-(r/r_0)^2}}
%\end{equation}
\begin{equation}
\label{eq:interactions2}
    V_{n}(r)=\alpha_0\biggr(\frac{\hbar^2}{2\mu_{\mathrm{2B}} r_0^2}\biggr)(r/r_0)^ne^{-(r/r_0)^2}
\end{equation}
\begin{equation}
\label{eq:interactions3}
    V(r)=\alpha_0\biggr(\frac{\hbar^2}{2\mu_{\mathrm{2B}} r_0^2}\biggr)e^{-(r/r_0)^2}+\alpha_1\biggr(\frac{\hbar^2}{2\mu_{\mathrm{2B}} r_1^2}\biggr)e^{-(r/r_1)^2}
\end{equation}
\end{subequations}
where $\alpha_0$ and $\alpha_1$ are dimensionless quantities that define the strength of the interaction, $r_0$, and $r_1$ give the range of the interaction, $\mu_{\mathrm{2B}}$ is the two--body reduced mass and $n$ defines some arbitrary power. For a given $r_0$, the $s$--wave scattering length and $p$--wave scattering volume is determined by varying the strength. In this work, the $s$--wave and $p$--wave interactions are tuned from non--interacting to the unitary limit, specifically the first poles in the scattering length and scattering volume, just before an $s$--wave or $p$--wave two--body bound state forms. Using these two--body interactions, three-- and four--interacting fermionic systems are studied near and at $s$--wave and $p$--wave unitary.

A Van der Waals interaction with a short--range cutoff is also considered in this work. The general form for this potential used with the ECG basis is given by the expression,
\begin{equation}
    \label{eq:vanderwaals}
    V(r)=-(C_q/r^q)(1-e^{-(r/r_0)^2})^p
\end{equation}
where $C_q$ is the coefficient of the potential tail at large $r$, $q$ is the power of the potential at large $r$, and $r_0$ and $p$ are parameters that control the short--range cutoff. The constraint on the value of $p$ is $2p\ge q$ to prevent the interaction from diverging at the origin. To easily evaluate matrix elements, the short--range factor $(1-e^{-(r/r_0)^2})^p$ is expanded using the binomial expansion. From Eq. \eqref{eq:vanderwaals}, the behavior of this interaction goes like $-C_q(1/r_0)^{2p}r^{(2p-q)}$ at small $r$ and the behavior goes as $-C_q/r^q$ at large $r$. The characteristic length scale for this potential is the standard definition $R_q=(1/2)[(2\mu/\hbar^2)C_q]^{1/(q-2)}$ in the limit as $r_0\rightarrow~0$. For the calculations involving Eq. \eqref{eq:vanderwaals}, the parameters used are $C_6=16$, $q=6$, $p=3~\mathrm{and}~p=4$, with $r_0$ changing depending on the scattering parameters of interest, namely the scattering volume for $p$--waves.

From the form of the two--body interactions given in Eqs. \eqref{eq:interactions} and \eqref{eq:vanderwaals}, a natural set of length and energy units are $\bar{r}_0$ and $E_0$, where $E_0=\hbar^2/2\mu_{2\mathrm{B}} \bar{r}_{0}^2$. With these definitions, the values of $r_0$ and $r_1$ are in units of $\bar{r}_0$ and the interaction $V(r)$ is given in units of $E_0$ by setting $\hbar=1$ and $m=1$, which is the convention used throughout this work for an equal--mass system. With the unit system defined in this way, the problem is recast in a more general way. For a given system, an appropriate choice of $\bar{r}_0$ and $E_0$ should be chosen. For example, in atomic systems with Van der Waals molecules, a good choice for $\bar{r}_{0}$ and $E_0$ would be the Van der Waals length ($r_{\mathrm{VdW}}$) and Van der Waals energy ($E_{\mathrm{VdW}}$). The Van der Waals length is defined as $r_{\mathrm{VdW}}=(1/2)[(2\mu_{\mathrm{2B}}/\hbar^2)C_6]^{1/4}$ and the Van der Waals energy is defined as $E_{\mathrm{VdW}}=\hbar^2/2\mu_{\mathrm{2B}}r_{\mathrm{VdW}}^2$. Likewise in nuclear systems, $\bar{r}_{0}$ and $E_{0}$ would have units of fm and MeV. The two--body interactions and their parameters used in this work are given in Table \ref{table:interaction_params}. The parameters listed are at the unitary limit for the $s$--wave (first pole in $a_{s}$) and $p$--wave (first pole in $V_{p}$) studies described in Secs. \ref{sec:section2} and \ref{sec:section3}, respectively.
%\textcolor{red}{Using the convention of $\hbar=1$ and $m=1$ described in the preceding paragraph, and from the form of the two--body interactions given in Eqs. \eqref{eq:interactions} and \eqref{eq:vanderwaals}, the units used throughout this work are Van der Waals units. Thus, the length scale is the Van der Waals length, defined as $r_{\mathrm{VdW}}=(1/2)[(2\mu_{\mathrm{2B}}/\hbar^2)C_6]^{1/4}$ and the energy scale is the Van der Waals energy given as $E_{\mathrm{VdW}}=\hbar^2/2\mu_{\mathrm{2B}}r_{\mathrm{VdW}}^2$. The two--body interactions and their parameters used in this work are given in Table \ref{table:interaction_params}. The parameters listed are at the unitary limit for the $s$--wave (first pole in $a_{s}$) and $p$--wave (first pole in $V_{p}$) studies described in Secs. \ref{sec:section2} and \ref{sec:section3}, respectively.}

\begin{table*}[ht]
\caption{Two--body interaction parameters used for calculations at the $s$--wave and $p$--wave unitary limits, where $a_{s}~\simeq~-1.0~\times~10^{12}~\bar{r}_{0}$ and $V_{p}~\simeq~-1.0~\times~10^{12}~\bar{r}_{0}^3$. The superscript $u$ indicates the quantities are given at the unitary limits.}

\label{table:interaction_params}
\begin{ruledtabular}
\begin{tabular}{cccccc}
    \multicolumn{6}{c}{\bf{$s$--wave ($\uparrow\downarrow$)}}\\
    \hline
    $\mathrm{Interaction}$ & $r_{0}(\bar{r}_{0})$ & $\alpha_{0}^{u}$ & $r_{1}(\bar{r}_{0})$ & $\alpha_{1}^{u}$ & $r_{\mathrm{eff}}(\bar{r}_{0})$\\
    \hline
    $V_{\eqref{eq:interactions2}},~n=0$ & $1.000000$ & $-2.684005$ & $\mathrm{N/A}$ & $\mathrm{N/A}$ & $1.435246$\\
    $V_{\eqref{eq:interactions2}},~n=1$ & $1.750000$ & $-2.853468$ & $\mathrm{N/A}$ & $\mathrm{N/A}$ & $3.013334$ \\
    $V_{\eqref{eq:interactions2}},~n=2$ & $2.000000$ & $-2.444496$ & $\mathrm{N/A}$ & $\mathrm{N/A}$ & $3.932112$ \\
    $V_{\eqref{eq:interactions2}},~n=0$ & $0.100000$ & $-2.684005$ & $\mathrm{N/A}$ & $\mathrm{N/A}$ & $0.143525$ \\
    \hline
    \multicolumn{6}{c}{\bf{$p$--wave ($\uparrow\uparrow$)}}\\
    \hline
        $\mathrm{Interaction}$ & $r_{0}(\bar{r}_{0})$ & $\alpha_{0}^{u}$ & $r_{1}(\bar{r}_{0})$ & $\alpha_{1}^{u}$ & $r_{\mathrm{eff}}(\bar{r}_{0}^{-1})$\\
    \hline
    $V_{\eqref{eq:interactions2}},~n=0$ & $1.000000$ & $-12.09931$ & $\mathrm{N/A}$ & $\mathrm{N/A}$ & $-2.058647$\\
    $V_{\eqref{eq:interactions2}},~n=1$ & $1.300000$ & $-12.14525$ & $\mathrm{N/A}$ & $\mathrm{N/A}$ & $-1.343086$\\
    $V_{\eqref{eq:interactions2}},~n=2$ & $2.000000$ & $-10.02293$ & $\mathrm{N/A}$ & $\mathrm{N/A}$ & $-0.772240$\\
    $V_{\eqref{eq:vanderwaals}},~q=6,~p=3$ & $0.995898$ & $16^{\footnote{\label{fn:vdw}For $V_{\eqref{eq:vanderwaals}}$, the strength parameter is the $C_{6}$ coefficient in Eq. \eqref{eq:vanderwaals} which is set to 16 in units of $E_{\mathrm{VdW}}r_{\mathrm{VdW}}^6$. In this case, $\bar{r}_{0}=r_{\mathrm{VdW}}$ and $E_0=E_{\mathrm{VdW}}$.}}$ & $\mathrm{N/A}$ & $\mathrm{N/A}$ & $-2.080183$\\
    $V_{\eqref{eq:vanderwaals}},~p=6,~q=4$ & $0.856721$ & $16^{\ref{fn:vdw}}$ & $\mathrm{N/A}$ & $\mathrm{N/A}$ & $-1.957820$\\
    $V_{\eqref{eq:interactions3}}^1$ & $1.000000$ & $26.66253$ & $2.000000$ & $-21.33003$ & $-0.838186$\\
    $V_{\eqref{eq:interactions3}}^2$ & $0.577350$ & $25.88748$ & $0.707107$ & $-34.94809$ & $-2.309161$\\
\end{tabular}
\end{ruledtabular}
\end{table*}

\section{Interacting fermions near the s--wave unitary limit\label{sec:section2}}
The first type of few--body systems looked at in this study are the fermionic systems interacting near the $s$--wave unitary limit, specifically at the first $s$--wave pole in $a_{s}$. This is accomplished by treating two cases; (1) treating the system with definite total angular momentum and parity $L^{\pi}$ and the spin of the individual particles separately, and (2) treating the system with a definite total angular momentum $L$ and total spin $S$. Case (1) is emphasised in this work, which is most relevant for experiment. We have computed hyperspherical potential curves for the three-- and four--body systems in different spin configurations and total orbital angular momentum using a single Gaussian two--body interaction. The strength of the two--body interaction was tuned to give $s$--wave scattering lengths in the range of $-\infty$ to 0. The $s$--wave scattering length dependence of the asymptotic $R^{-3}$ coefficient near $s$--wave unitarity was computed for each symmetry as well as the effective angular momentum $l_{\mathrm{eff}}$ controlling the $R^{-2}$ coefficient at unitarity, where at large $R$, $U_{\nu}(R)\rightarrow \hbar^2l_{\mathrm{eff}}(l_{\mathrm{eff}}+1)/2\mu R^2$. The primary results reported in this section are for a single Gaussian two--body interaction with an interaction range $r_0$ (see Eq. \eqref{eq:interactions}).

\subsection{Three and four 2-component fermions at the s--wave unitary limit}
\begin{figure}[!ht]
    \centering
    \subfigure[]{\includegraphics[width=8.6 cm]{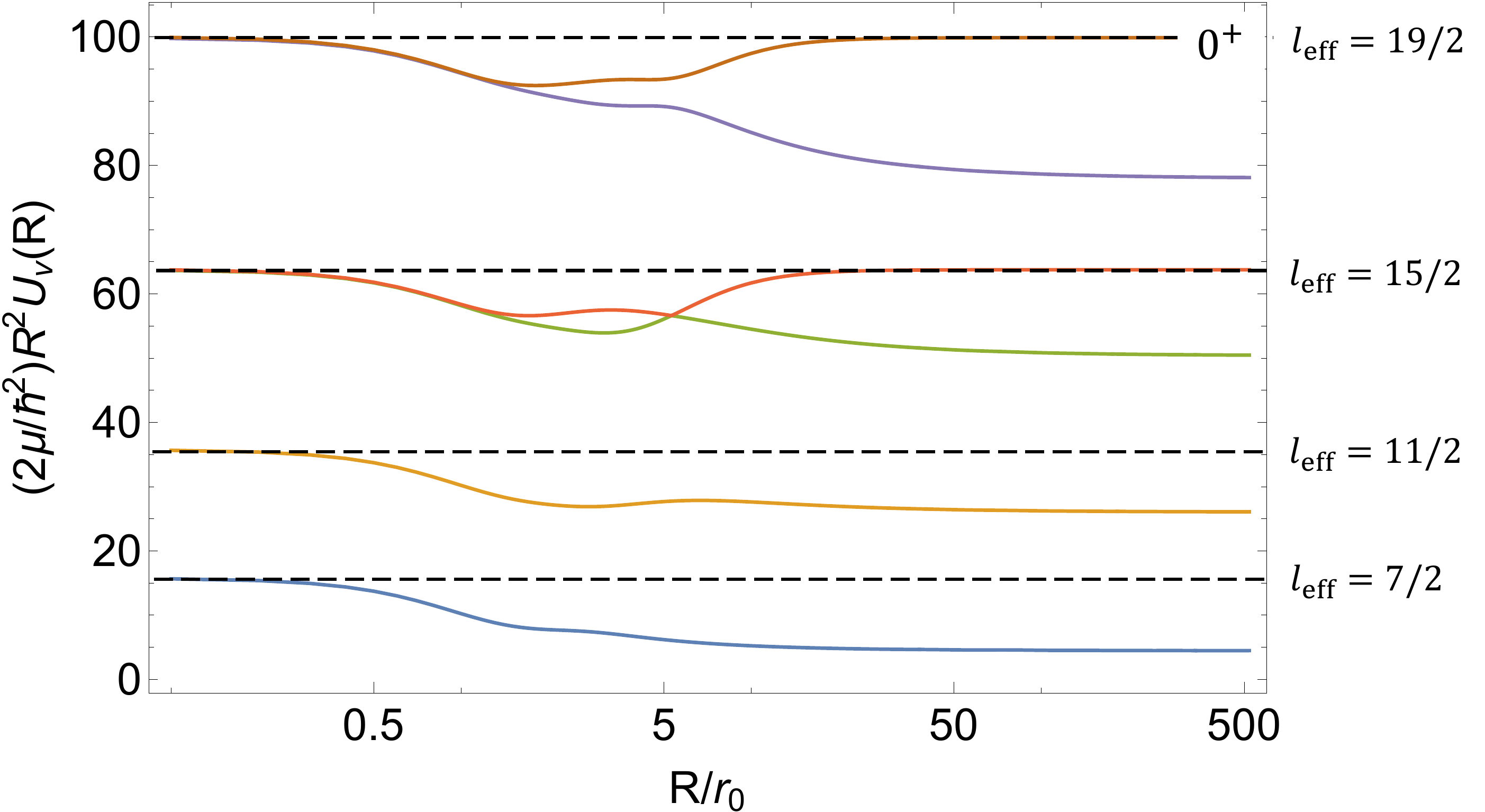}}

    \subfigure[]{\includegraphics[width=8.6 cm]{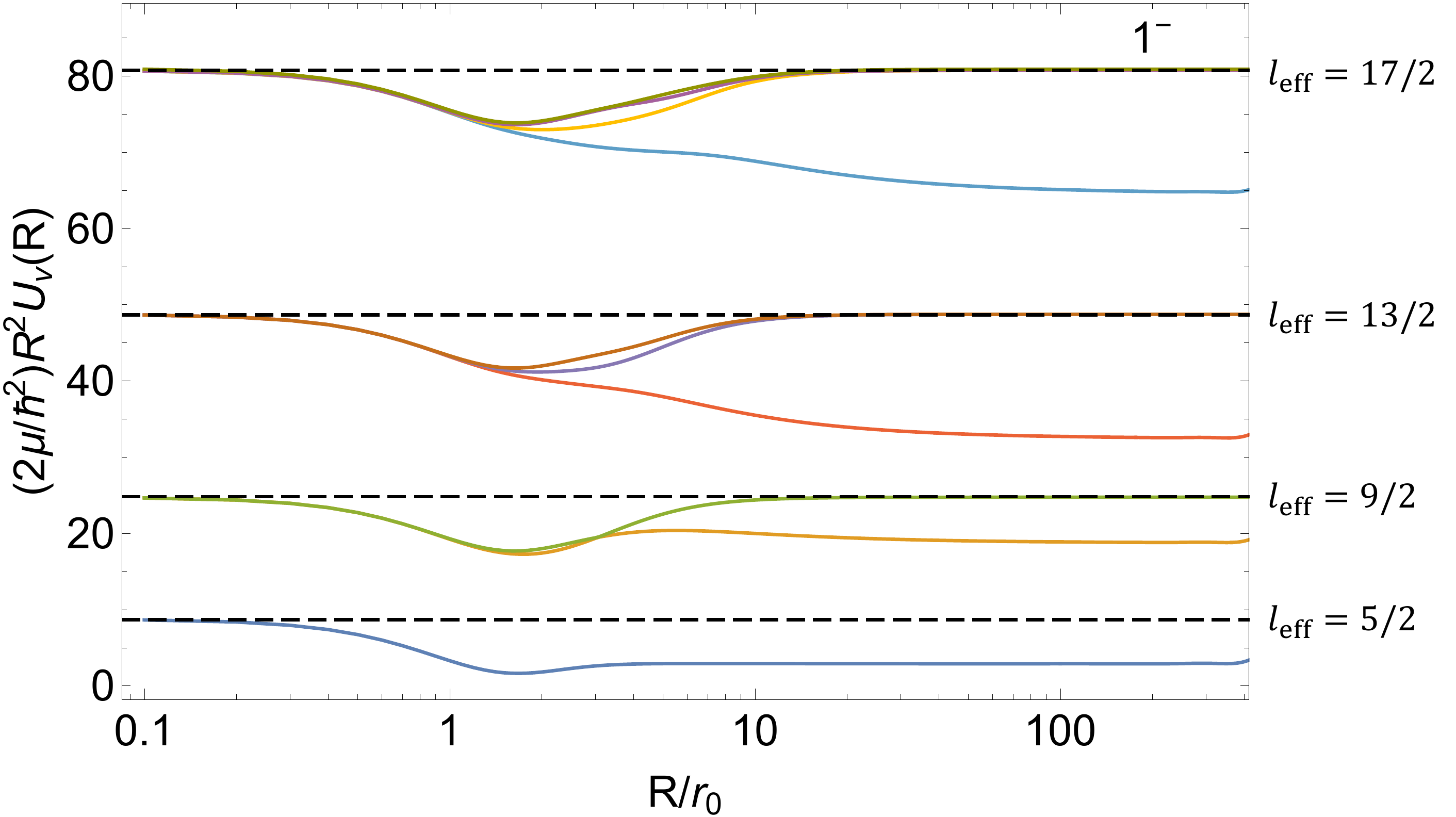}}
    
    \subfigure[]{\includegraphics[width=8.6 cm]{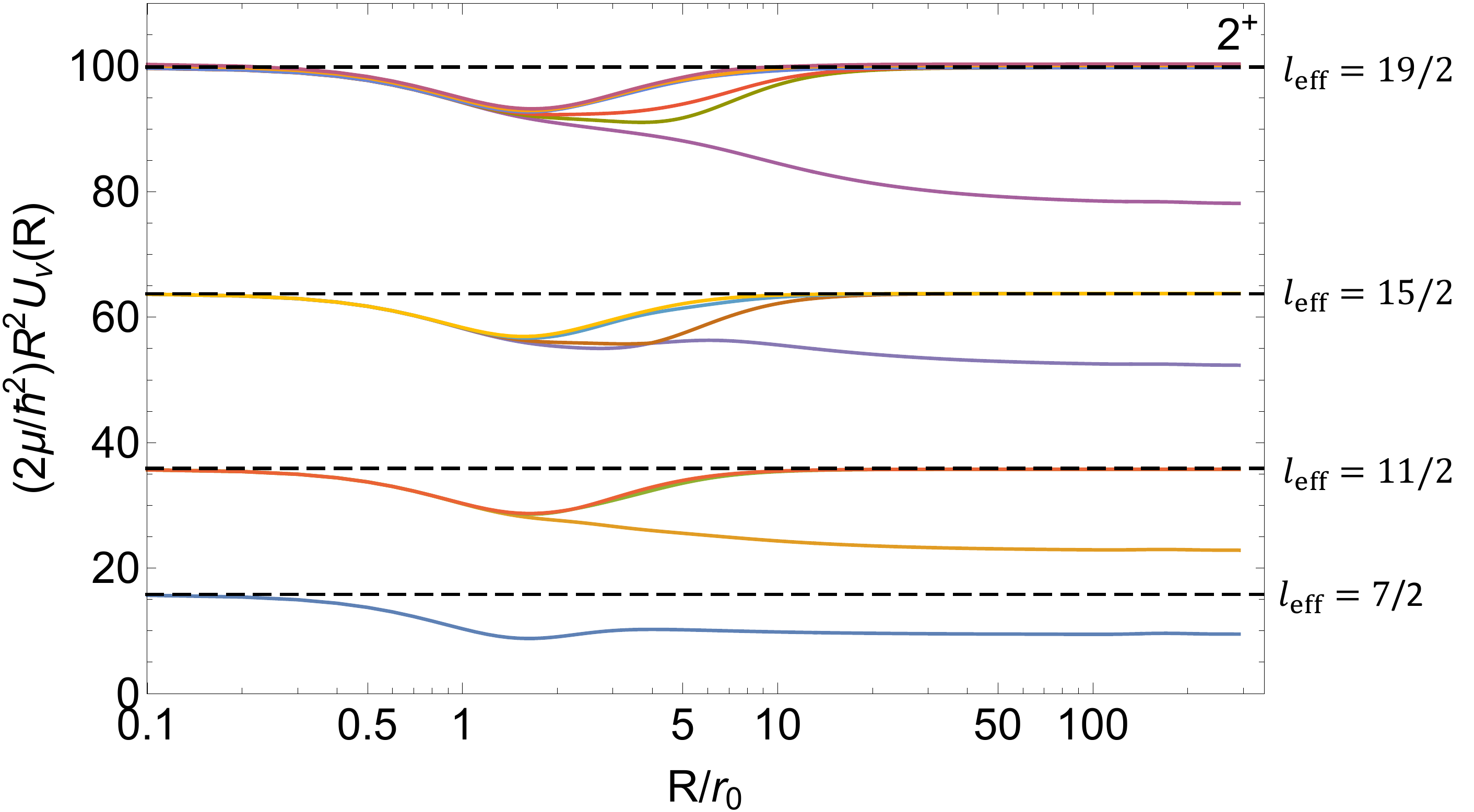}}
    
    \caption{\label{fig:3_Body_rsqU} The lowest few Born--Oppenheimer potential curves for the $(\uparrow\uparrow\downarrow)$ equal mass three--body system are shown for the symmetries $L^{\pi}$=$0^{+}$ in (a), $1^{-}$ in (b) and $2^{+}$ in (c). The two--body interactions between all particles are at the first $s$--wave unitary limit before the formation of an $s$--wave dimer. The horizontal dashed lines represent the non--interacting rescaled potentials with their effective angular momentum quantum numbers $l_{\mathrm{eff}}$ labeled to the right.} The hyperradius has been rescaled by the range of the Gaussian interaction $r_0$ and shown on a log scale. For some of the higher degenerate channels, the potentials deviate at small hyperradius due to incomplete basis set convergence.
\end{figure}

For the first case, the three--body and four--body cases of equal mass interacting particles are studied in different spin configurations, i.e. $(\uparrow\uparrow\downarrow)$, $(\uparrow\uparrow\downarrow\downarrow)$, and $(\uparrow\uparrow\uparrow\downarrow)$. The $(\uparrow\uparrow\uparrow\uparrow)$ spin configuration is reserved for Sec. \ref{sec:section3} that treats $p$--wave interactions. The two--body interaction between each pair of opposite spin  particles are set to be the same and tuned over a range of large $s$--wave scattering lengths. Gaussian interactions are used in this analysis, with an interaction range $r_0$ and strength $\alpha_0$, represented by Eq. \eqref{eq:interactions}. The strength $\alpha_0$ is tuned to give different $s$--wave scattering lengths in the range $0<|a_s|<\infty$. This work focuses on the region where $a_s$ is negative up to the first pole in $a_s$. The three--body hyperradial potentials at the $s$--wave unitary limit are shown in Figure \ref{fig:3_Body_rsqU} for $L^{\pi}=0^+,1^-$, and $2^+$ in the $(\uparrow\uparrow\downarrow)$ spin configuration.

Figure \ref{fig:3_Body_rsqU} shows the lowest few $N=3$ hyperradial potential curves for the following symmetries at the $s$--wave unitary limit: $0^{+}$ in $(a)$, $1^{-}$ in $(b)$, and $2^{+}$ symmetry in $(c)$. These potentials shown have been rescaled by $(2\mu/\hbar^2)R^2$ and are shown on a log scale in $R/r_0$. In this representation, the curves approach a constant value at large values of $R/r_0$, which represents the effective angular momentum barrier which equals the value of $l_{\mathrm{eff}}(l_{\mathrm{eff}}+1)$, where $l_{\mathrm{eff}}$ is the effective angular momentum quantum number. In the three--body case, the $l_{\mathrm{eff}}$ takes on half--integer values for the non--interacting case (i.e. the lowest $l_{\mathrm{eff}}$ for the symmetries considered here are $7/2$, $5/2$, and $7/2$ for $0^+$, $1^-$, and $2^+$, respectively).  When the scattering length and scattering volume are both finite, these same noninteracting values of $l_{\mathrm{eff}}$ still apply at large $R$. At unitarity, however, the value of $l_{\mathrm{eff}}$ gets modified to a lower value in some of the channels, indicated by the lower deviations at large $R/r_0$ in Fig. \ref{fig:3_Body_rsqU}. The reduction of the effective angular momentum barrier has been extensively studied over the years, most notably by Werner and Castin \cite{PhysRevLett.97.150401} for the three boson and fermion case, and by Blume and coworkers for the four fermion case \cite{RakshitBlume2012pra,vonStecher2007prl}.  This reduction of the coefficient of $R^{-2}$ in the potential curve at $R\rightarrow \infty$ appears to be the same mechanism that reduces the analogous coefficient at unitarity for a system of 3 bosonic (or different spin fermionic) particles in the famous Efimov effect.

\begin{figure}[H]
    \centering
    \subfigure[]{\includegraphics[width=8.6 cm]{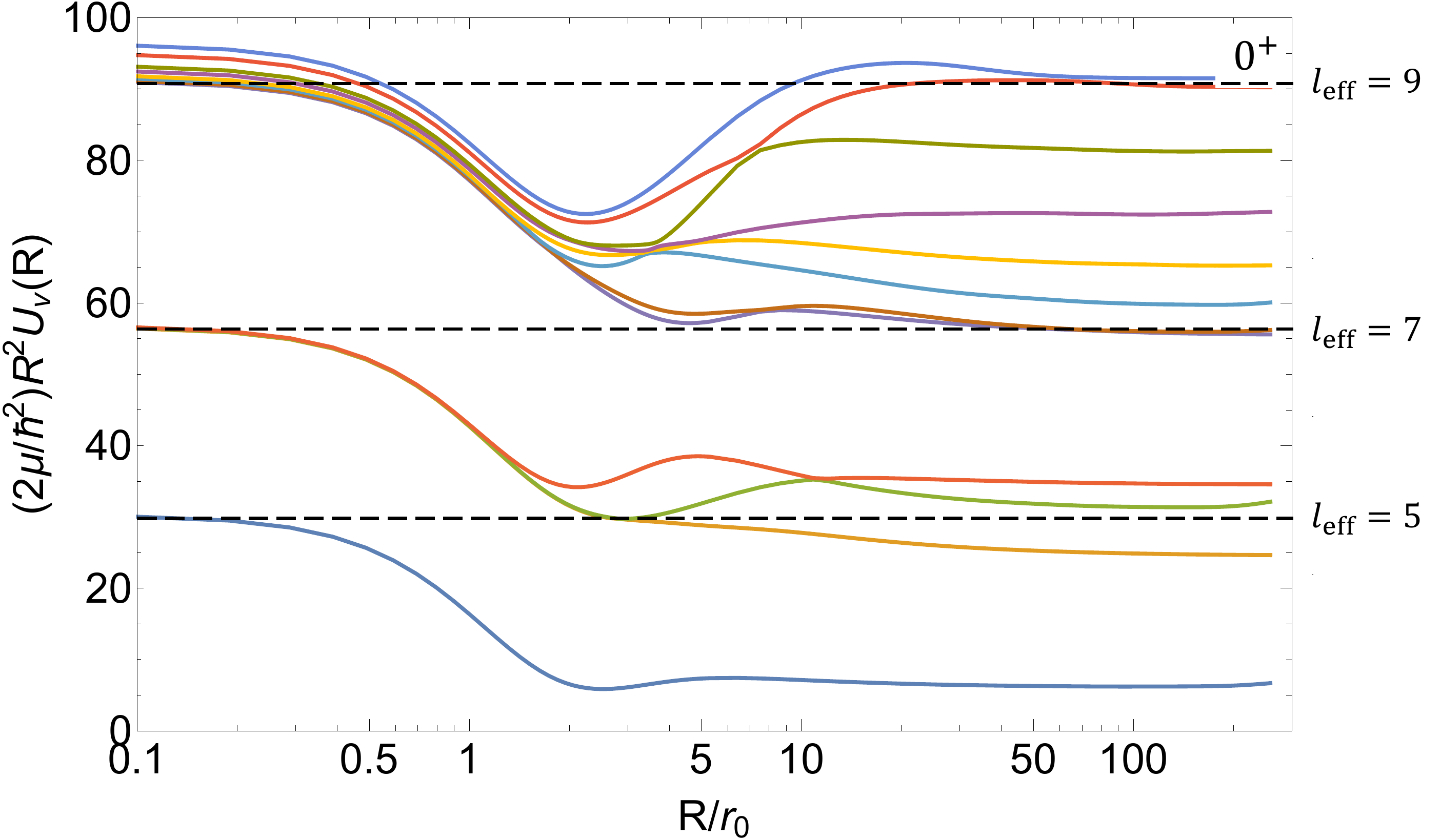}}

    \subfigure[]{\includegraphics[width=8.6 cm]{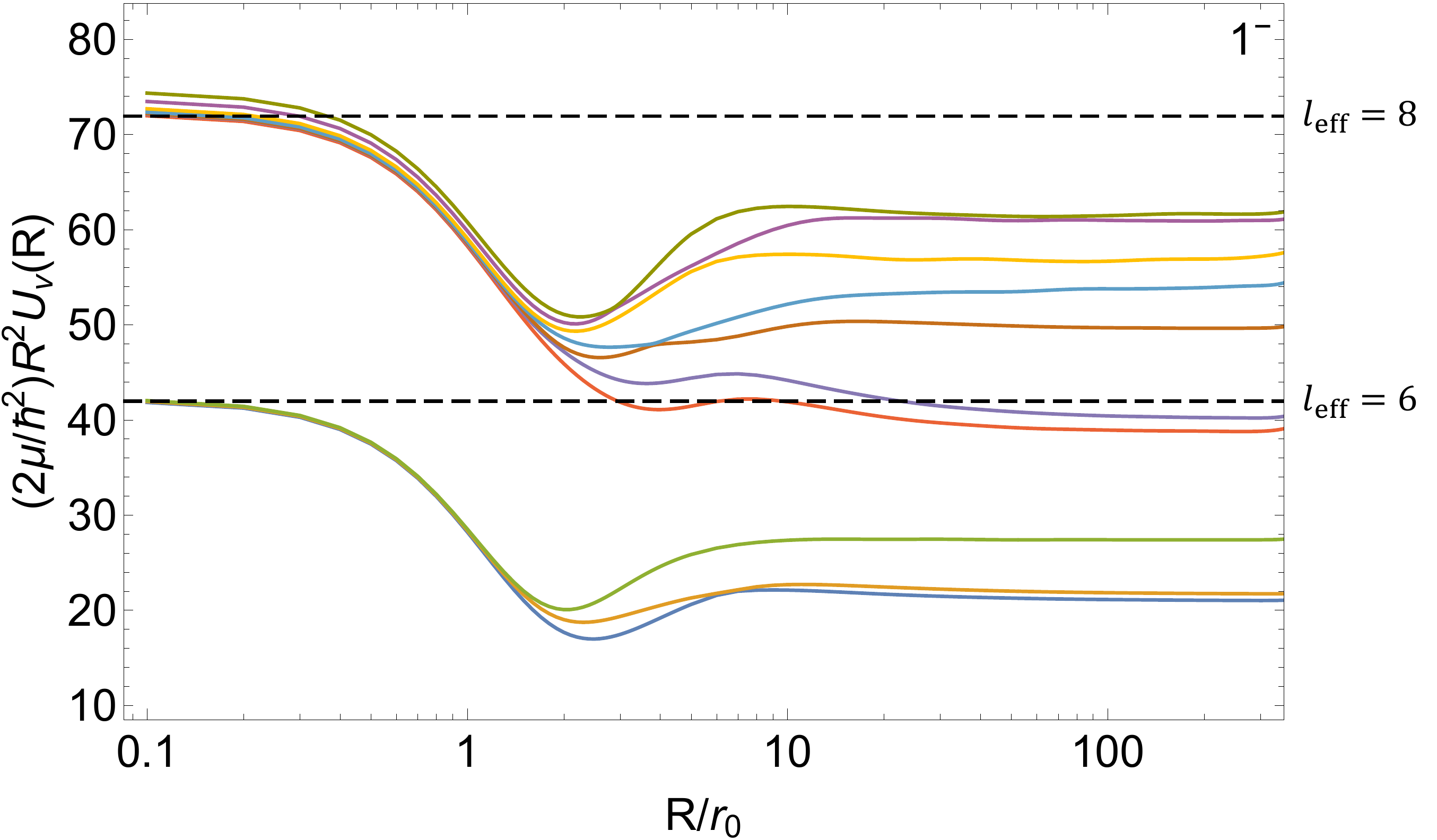}}
    
    \subfigure[]{\includegraphics[width=8.6 cm]{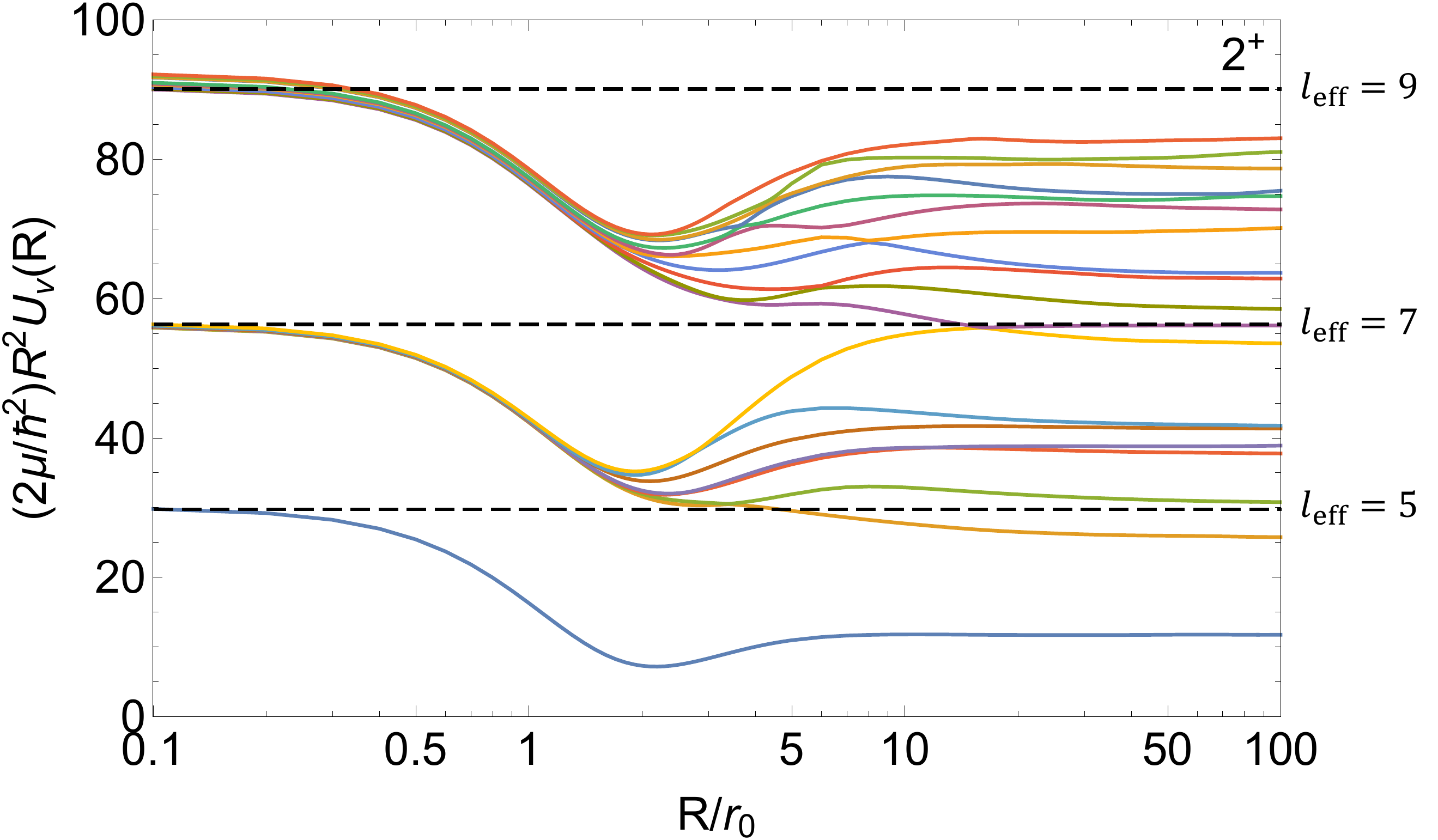}}
    
    \caption{\label{fig:4_Body_rsqU_uudd} The lowest few Born--Oppenheimer potential curves for the $(\uparrow\uparrow\downarrow\downarrow)$ equal mass four--body system are shown for the symmetries $L^{\pi}$=$0^{+}$ in (a), $1^{-}$ in (b) and $2^{+}$ in (c). The strength of the two--body interaction between all particles are scaled give an infinite $s$--wave scattering length. The horizontal dashed lines represent the non--interacting rescaled potentials with their effective angular momentum quantum numbers $l_{\mathrm{eff}}$ labeled to the right. The hyperradius has been rescaled by the range of the Gaussian interaction $r_0$. For some of the higher degenerate channels, the potentials deviate at small hyperradius due to incomplete basis set convergence.}
\end{figure}

\begin{figure}[H]
    \centering
    \subfigure[]{\includegraphics[width=8.6 cm]{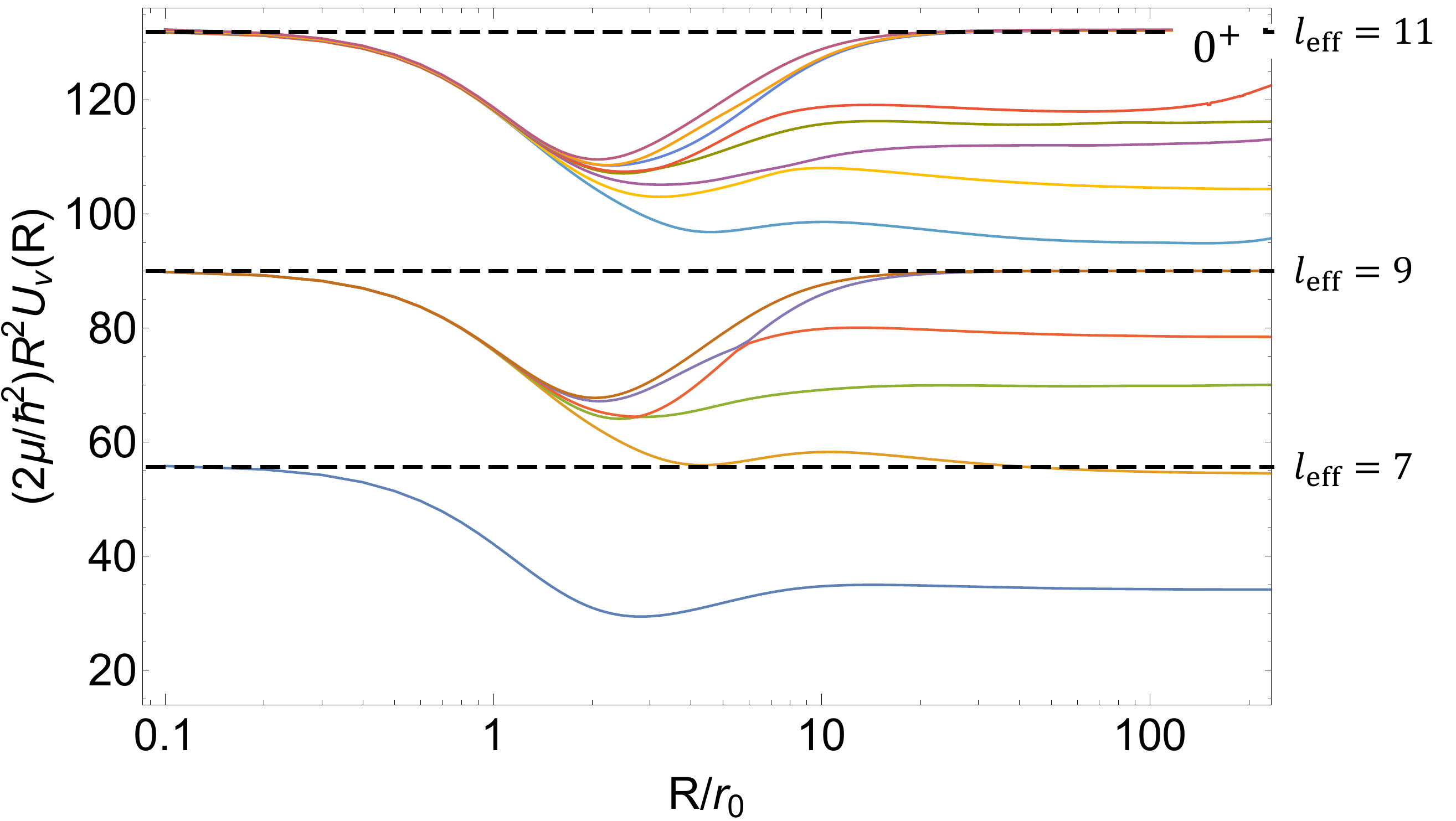}}

    \subfigure[]{\includegraphics[width=8.6 cm]{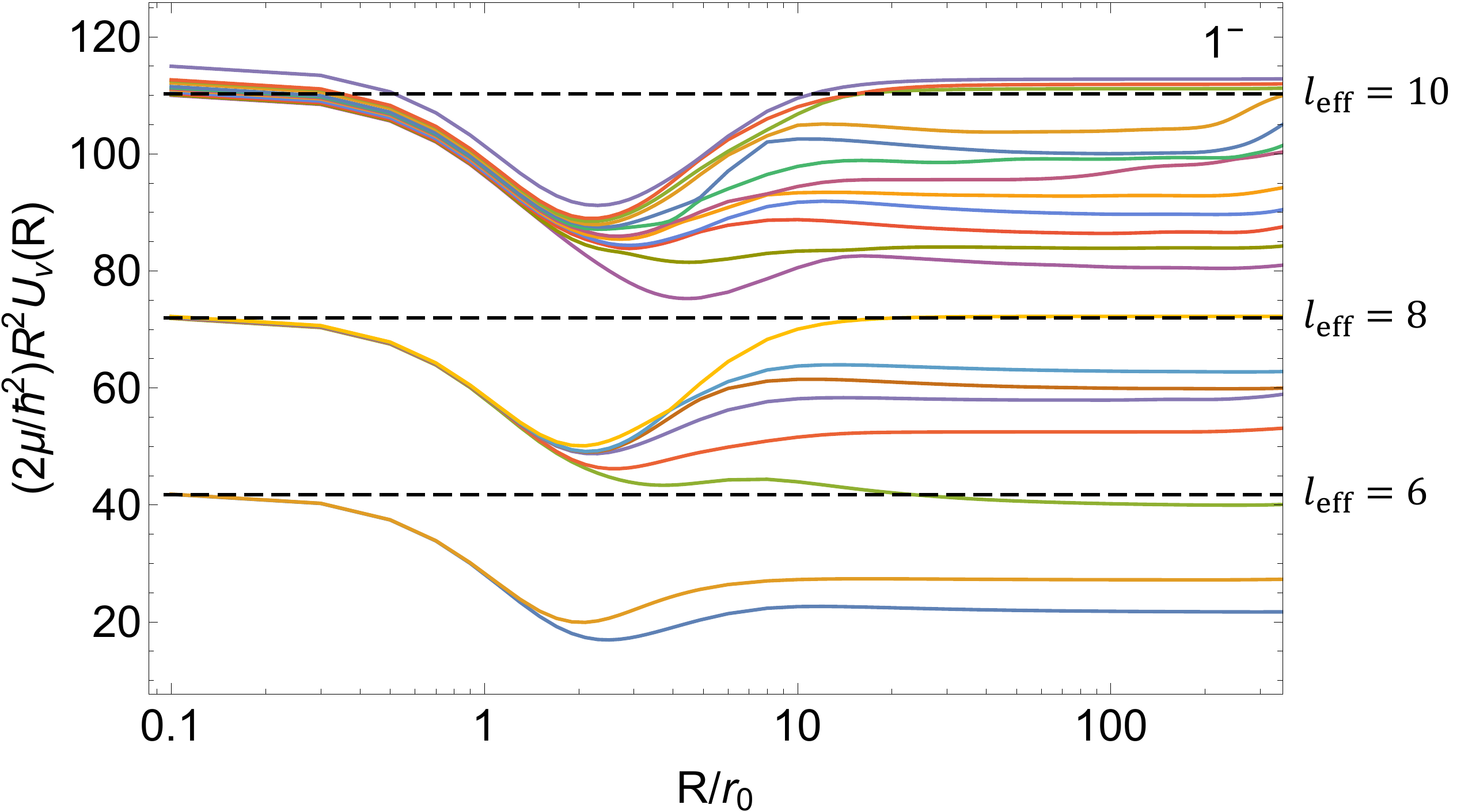}}
    
    \subfigure[]{\includegraphics[width=8.6 cm]{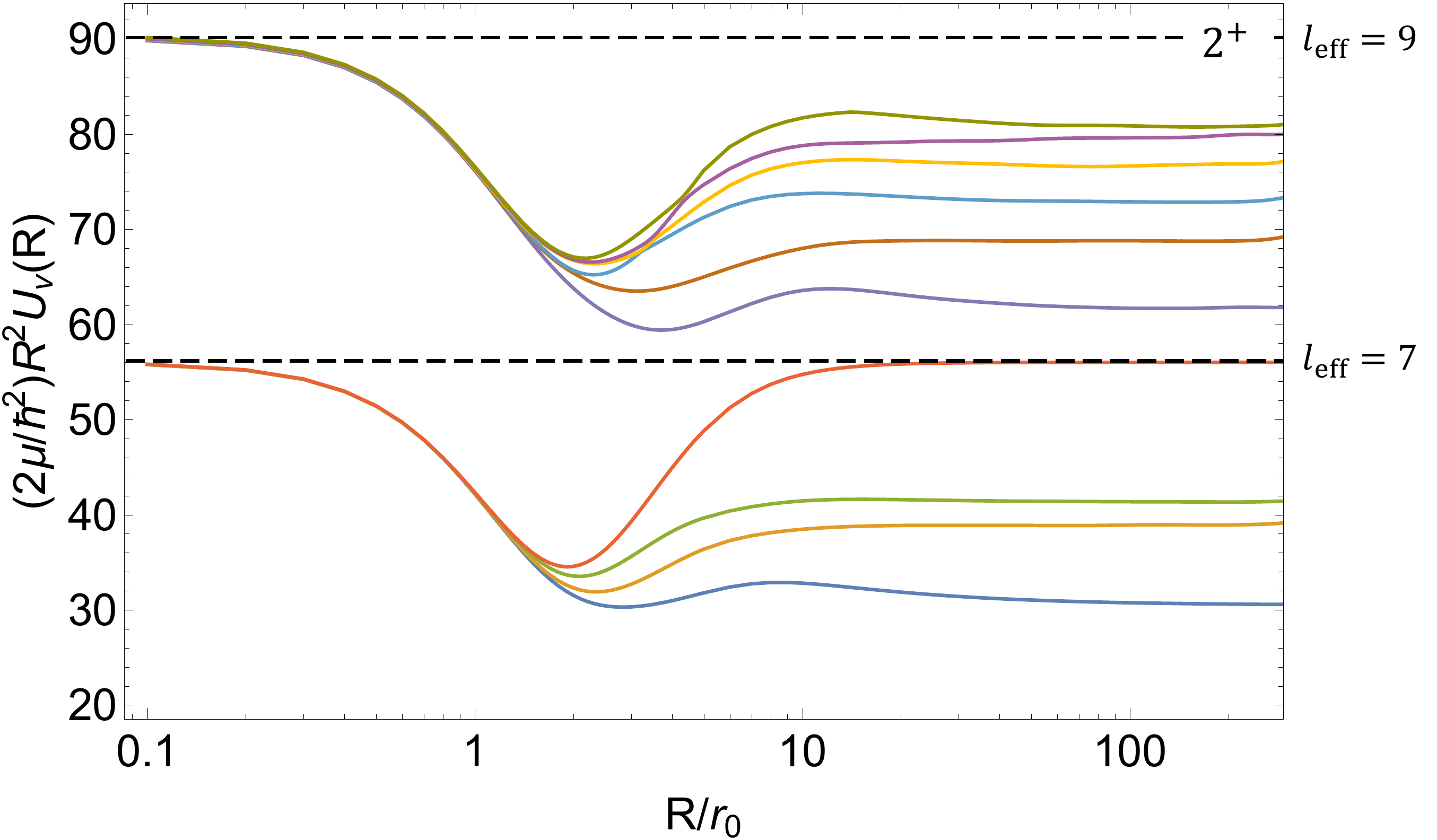}}
    
    \caption{\label{fig:4_Body_rsqU_uuud} The lowest few Born--Oppenheimer potential curves for the $(\uparrow\uparrow\uparrow\downarrow)$ equal mass four--body system are shown for the symmetries $L^{\pi}$=$0^{+}$ in (a), $1^{-}$ in (b) and $2^{+}$ in (c). The strength of the two--body interaction between all particles are scaled give an infinite $s$--wave scattering length. The horizontal dashed lines represent the non--interacting rescaled potentials with their effective angular momentum quantum numbers $l_{\mathrm{eff}}$ labeled to the right. The hyperradius has been rescaled by the range of the Gaussian interaction $r_0$. For some of the higher degenerate channels, the potentials deviate at small hyperradius due to incomplete basis set convergence.}
\end{figure}

For the four--body case, Figures \ref{fig:4_Body_rsqU_uudd} and \ref{fig:4_Body_rsqU_uuud} show plots similar to the three--body case discussed previously. These figures display the $0^+$, $1^-$, and $2^+$ symmetries for the $(\uparrow\uparrow\downarrow\downarrow)$ and $(\uparrow\uparrow\uparrow\downarrow)$ spin configurations, respectively. Like in Fig. \ref{fig:3_Body_rsqU}, these plots show a few of the Born--Oppenheimer potentials plotted versus $R/r_0$ on a log scale that highlights the effect of unitarity-limited two--body $s$--wave interactions on the effective angular momentum barrier in the hyperradial equation. For the $(\uparrow\uparrow\downarrow\downarrow)$ spin configuration, the lowest effective angular momentum quantum numbers in the non--interacting limit are $l_{\mathrm{eff}}=5$, $6$, and $5$ for the $0^+$, $1^-$, and $2^+$ symmetries, respectively. In the $(\uparrow\uparrow\uparrow\downarrow)$ spin configuration, the lowest effective angular momentum quantum numbers are respectively $l_{\mathrm{eff.}}=7$, $6$, and $7$ for the $0^+$, $1^-$, and $2^+$ symmetries. For both symmetries, it should be noted that the lowest non--interacting value of $l_{\mathrm{eff.}}$ is for the $1^-$ symmetry among the ones presented.  Since the generalized Wigner threshold law for a squared 4-body recombination scattering matrix element is proportional to $k^{2l_{\mathrm{eff}}+1}$, one expects the $1^-$ symmetry to be the dominant recombination symmetry out of these 3 parity-favored cases of $(\uparrow\uparrow\uparrow\downarrow)$. In both the three--body and four--body systems, we focus on the hyperradial potentials that go to the modified centrifugal-type barrier at large hyperradius when the particles interact near the unitary limit, where they exhibit universal behavior when tuning the scattering length for $|a_s/r_0|>10$. This universal behavior manifests in a long--range $1/R^3$ term that depends only on the $s$--wave scattering length and the symmetry of the system.

\subsection{Universal behavior of the hyperradial potentials in the $N$--body continuum}
When investigating the $s$--wave behavior in $N$--body systems, an interesting behavior arises in the hyperradial potentials representing the continuum states. Near the unitary regime, the long--range form of some of the continuum channels exhibits a $1/R^3$ behavior. The form of the Born--Oppenheimer potentials representing the $N$--body continuum at large $R$ is given by
\begin{equation}
    \label{eq:longrangeform}
    U_{\nu}^N(R)\rightarrow\frac{\hbar^2}{2\mu}\biggr[\frac{l_\mathrm{eff}(l_\mathrm{eff}+1)}{R^2}+\frac{C_{3,\nu}^N}{R^3}\biggr]
\end{equation}
where $l_{\mathrm{eff}}$ is the effective angular momentum quantum number, the coefficient $C_{3,\nu}^N$ is defined as $C_{3,\nu}^N=\mathcal{C}_{\nu}^N a_s$, where $a_s$ is the two--body $s$--wave scattering length and $\mathcal{C}_{\nu}^N$ is a universal $N$--body coefficient that depends on the particle statistics and other quantum numbers of the system. The adiabatic potentials $W_{\nu}(R)$ that includes the diagonal non--adiabatic correction $Q_{\nu\nu}(R)$, also follow this same long--range behavior, since the diagonal non--adiabatic correction falls--off faster than $1/R^3$ . This universal long--range behavior presents itself in the hyperspherical channels whose effective angular momentum quantum number is modified from the non--interacting value at infinite scattering length. The linear dependence of the $1/R^3$ term on the scattering length has been shown for these hyperradial potentials in the continuum for the three--boson system at large scattering length \cite{PhysRevA.82.022706,ColussiV}. As is shown in what follows, the hyperspherical potentials that approach the non--interacting limit at large $R$ for infinite scattering length do so on a shorter length scale than the potentials that approach the reduced $1/R^2$ coefficient. The coefficient $\mathcal{C}_{\nu}^N$ is computed for different total angular momentum states of a given $L$ for various spin configurations, focusing on the three and four identical particle systems with $L^{\pi}=[0^+, 1^{-}, 2^{+}]$ in this work. 

The spin configurations considered are some total spin states denoted by the total spin quantum number $S$, and also different individual spin configurations, specifically the $(\uparrow \uparrow \downarrow)$ for the three--body system, and likewise the $(\uparrow \uparrow \downarrow \downarrow)$ and $(\uparrow \uparrow \uparrow \downarrow)$ configurations for the four particle system. For the treatment of identical particles, the $s$--wave interaction between the $(\uparrow \downarrow)$ two--body system is tuned from non--interacting to the first $s$--wave unitary limit, where a low--energy $s$--wave dimer forms. In some symmetries, tuning the $(\uparrow\uparrow)$ two--body $p$--wave interaction strength will lead to the formation of a trimer state in the $(\uparrow \uparrow \downarrow)$ configuration.

\begin{figure}[H]
    \centering
    \includegraphics[width=8.6 cm]{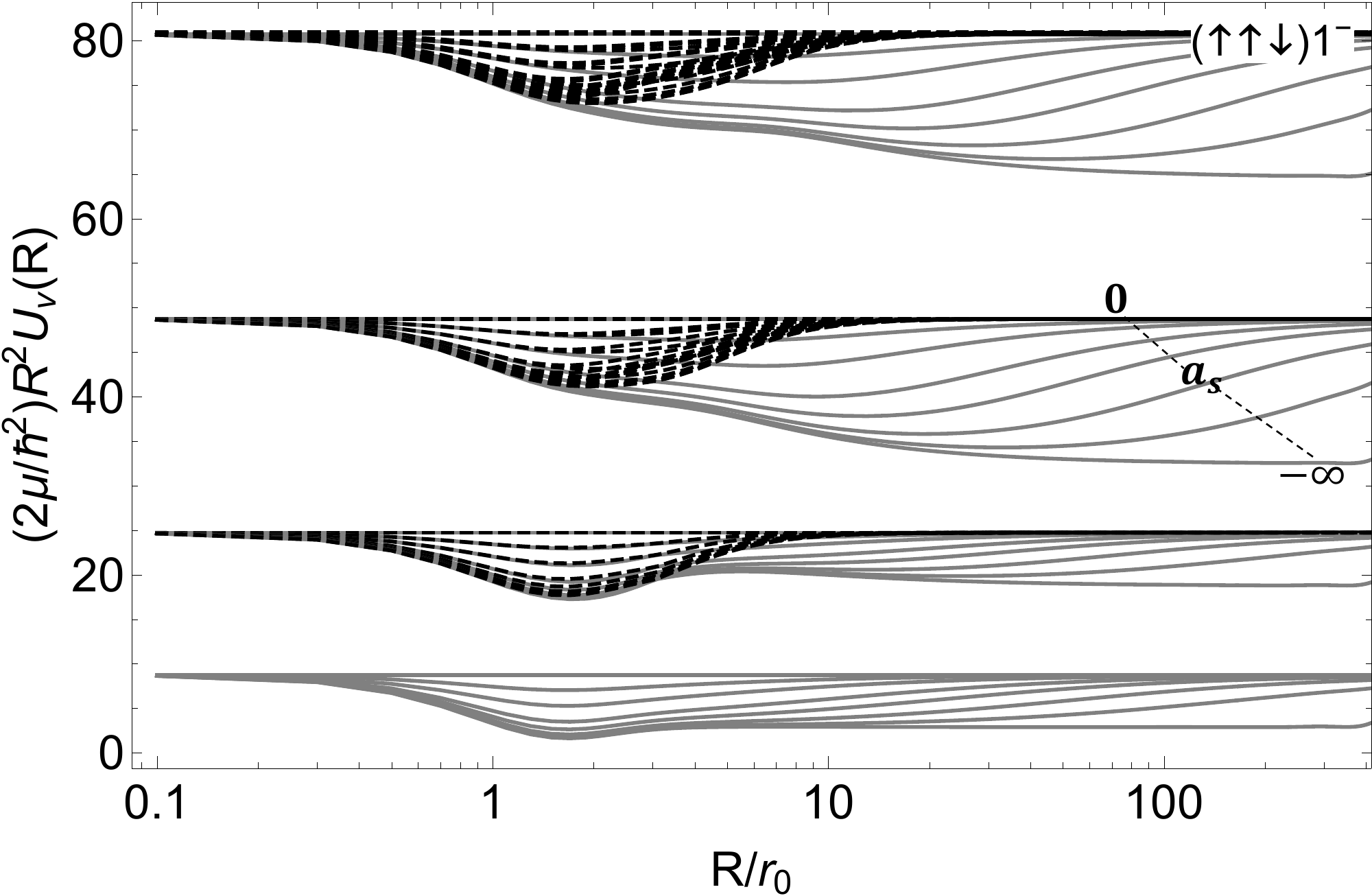}
    
    \caption{\label{fig:3_Body_rsqU_as} The lowest few Born--Oppenheimer potential curves for the $(\uparrow\uparrow\downarrow)$ equal mass three--body system are shown for the $1^{-}$ symmetry. The solid curves correspond to potentials that exhibit a reduced value of $l_{\mathrm{eff}}$ at the unitary limit for large hyperradius, whereas the dashed potentials go to the non--interacting potentials at large hyperradius. The strength of the two--body interactions between all particles are rescaled from the non--interacting limit up to the $s$--wave unitary limit ($a_s\rightarrow-\infty$). For each set of curves representing a different channel $\nu$, an increase in scattering length on the negative side corresponds to a curve in the set. Reading from highest to lowest, the highest curve is the non--interacting potential, the lowest is the hyperradial potential at the $s$--wave unitary limit, and a potential in between is for a finite scattering length. The hyperradius has been rescaled by the range of the Gaussian interaction $r_0$. The structure of the potential curves for different systems and symmetries are qualitatively similar to the potentials shown here, thus are not shown.}
\end{figure}

In this section, the long--range behavior of the lowest few hyperradial potentials in each spin configuration and system size is characterized. Starting with the three--body case for the $(\uparrow\uparrow\downarrow)$ spin configuration, some of the lowest hyperradial potentials have a modified angular momentum barrier at the $s$--wave unitary limit for all symmetries, which is clearly represented in Fig. \ref{fig:3_Body_rsqU}. For these hyperradial potentials, the long--range behavior of the potential for finite scattering length close to unitarity has the form of Eq. \eqref{eq:longrangeform}. To obtain the symmetry dependent coefficient $\mathcal{C}_{\nu}$, an analysis on the hyperradial potentials are performed at different scattering lengths. In Figure \ref{fig:3_Body_rsqU_as}, a subset of hyperradial potentials are shown to illustrate the structure of the long--range behavior at different scattering lengths. The hyperradial potentials for the other systems and symmetries have the same qualitative structure and are not shown here.

\begin{figure}[H]
    \centering
    \subfigure[]{\includegraphics[width=8.6 cm]{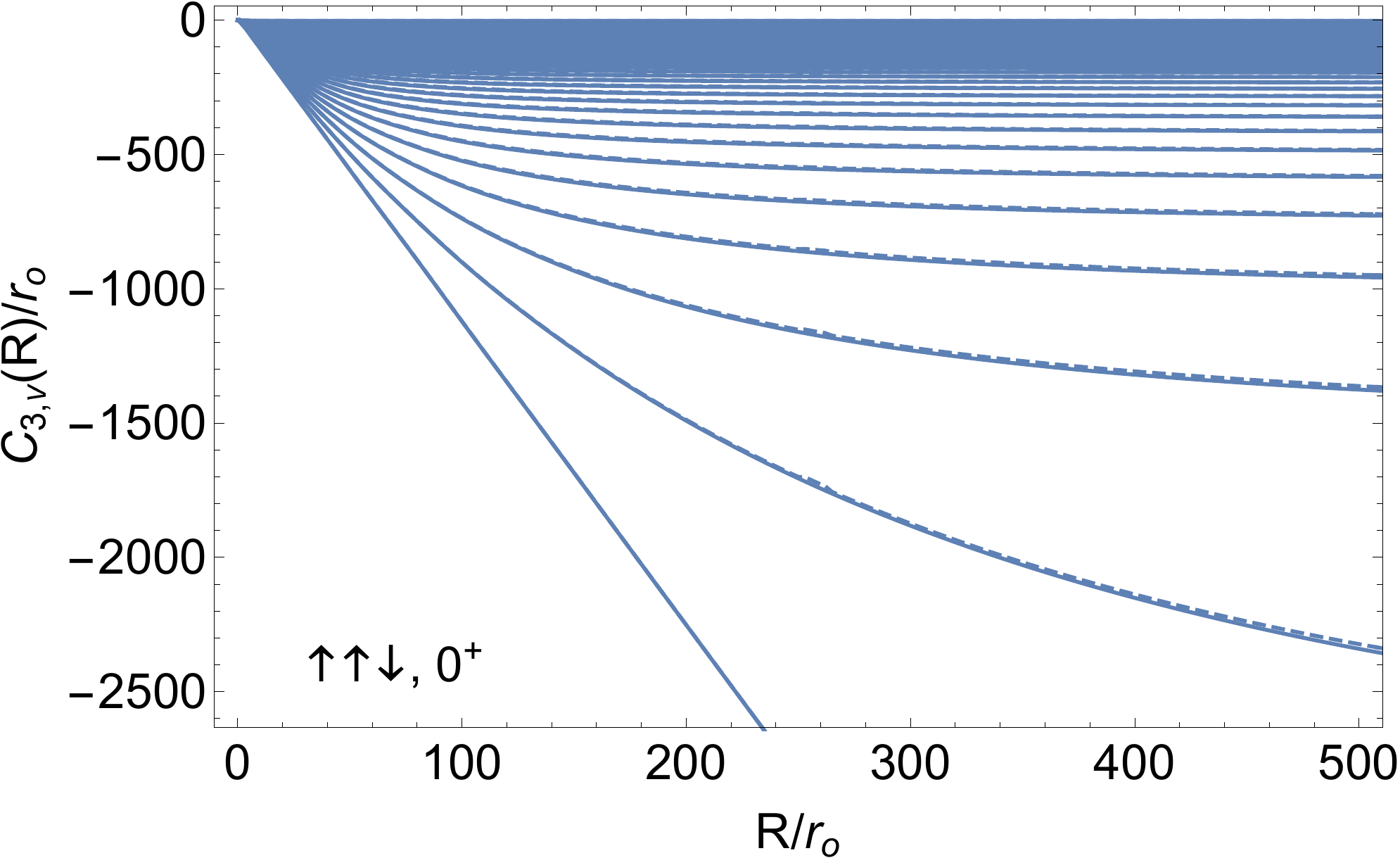}}

    \subfigure[]{\includegraphics[width=8.6 cm]{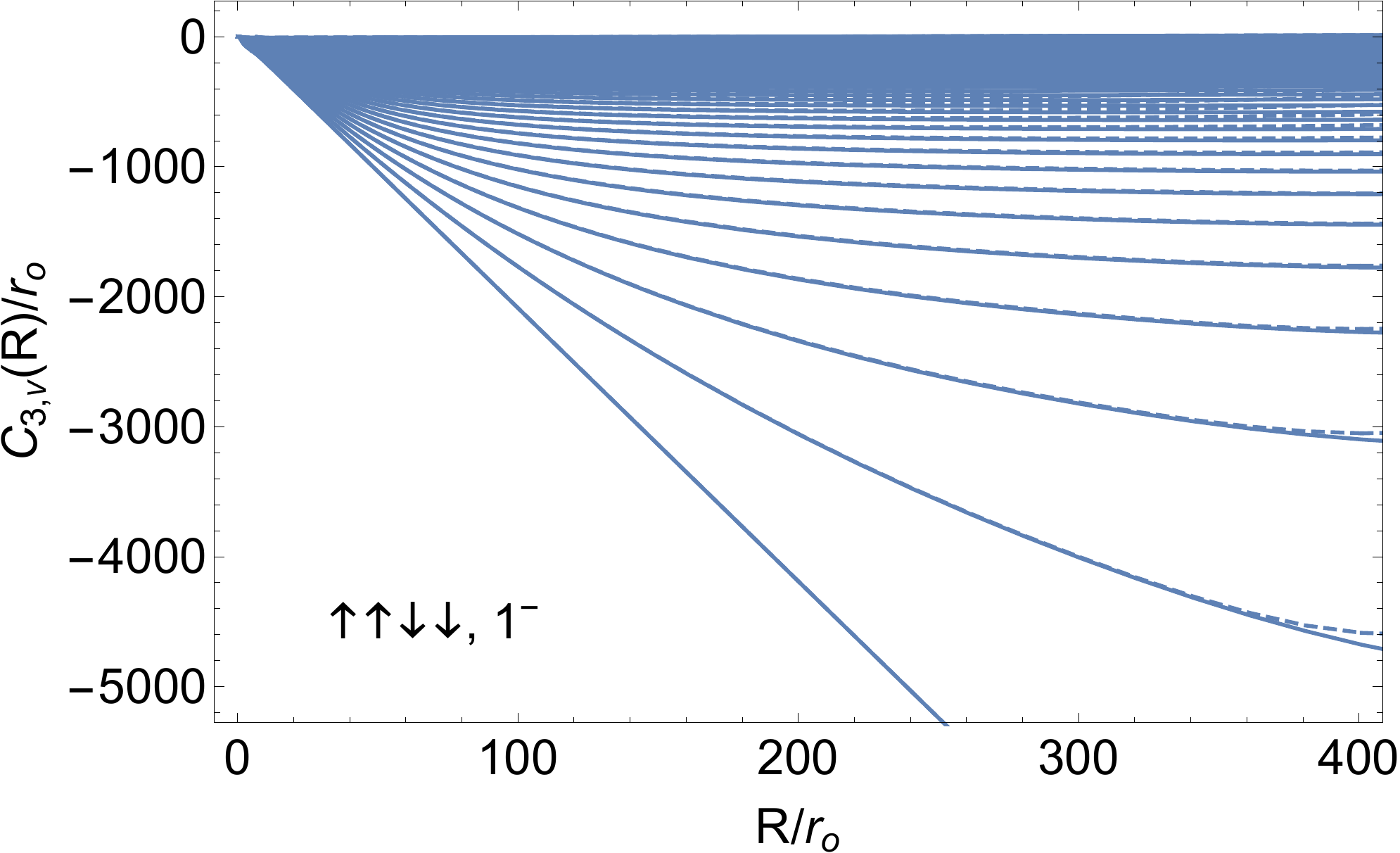}}
    
    \subfigure[]{\includegraphics[width=8.6 cm]{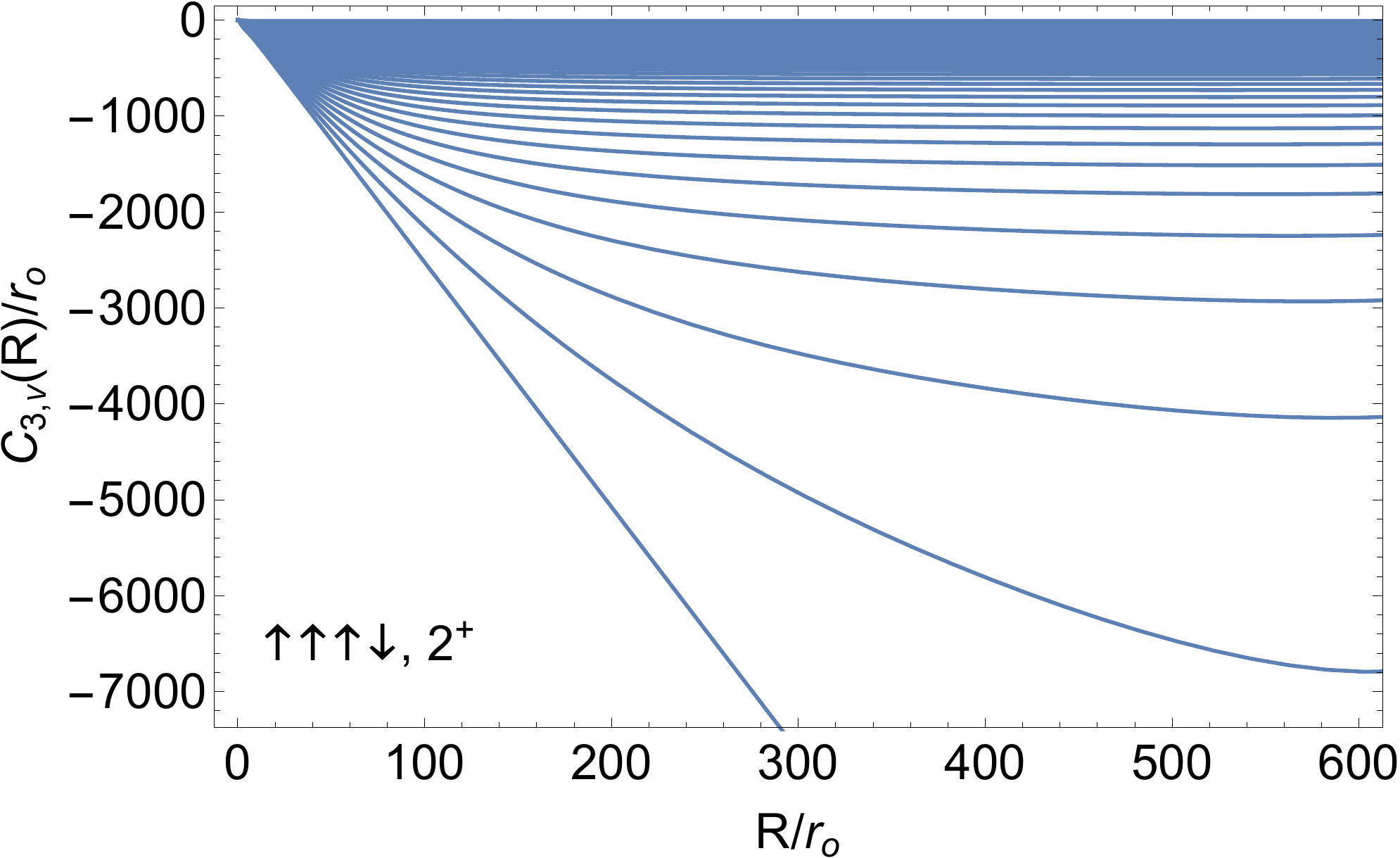}}
    
    \caption{\label{fig:4_Body_C3} The lowest few Born--Oppenheimer potential curves for the $(\uparrow\uparrow\downarrow)$ equal mass three--body system for the $0^{+}$ symmetry in (a), the $(\uparrow\uparrow\downarrow\downarrow)$ equal mass four--body system for the $1^{-}$ symmetry in (b) and the $(\uparrow\uparrow\uparrow\downarrow)$ equal mass four--body system for the $2^{+}$ symmetry in (c). The strength of the two--body interactions between all particles are rescaled from the non--interacting limit up to the $s$--wave unitary limit ($a_s\rightarrow-\infty$). The solid curves are for the adiabatic potential and the dashed curves include the second--derivative non--adiabatic correction. The hyperradius has been rescaled by the range of the Gaussian interaction $r_0$. The curves for the other symmetries are qualitatively similar and not shown here.}
\end{figure}

In Fig. \ref{fig:3_Body_rsqU_as}, the lowest few Born--Oppenheimer potentials are shown to highlight key features that arise when tuning the two--body scattering length from non--interacting to unitarity. As the $s$--wave scattering length increases on the negative side, some of the potentials start to exhibit a long--range deviation from the non--interacting limit as the angular momentum barrier transitions from the non--interacting value (finite scattering length) to the modified value (infinite scattering length) at large hyperradius. When the hyperradial potentials do not exhibit a modified barrier at infinite scattering length (dashed curves), the deviation from the non--interacting potential appears to be short--range. This is visible in Fig. \ref{fig:3_Body_rsqU_as} where some of the potentials reach non--interacting behavior already for $R/r_0\sim10-12$. The exact form of the behavior of these potentials (dashed lines) as they start deviating from the non--interacting limit at smaller hyperradius is not studied here, where the focus is on the potentials that go to the reduced barrier at the unitary limit. For the hyperradial potentials that go to the modified barrier at infinite scattering length (solid curves), the long--range deviation can be represented graphically by plotting $C_{3,\nu}(R)=R[(2\mu/\hbar^2)R^2U_{\nu}(R)-l_{\nu}(l_{\nu}+1)]$, which is based on Eq. \eqref{eq:longrangeform}. Plots of the function $C_{3,\nu}(R)$ for the lowest hyperradial potential for the three symmetries in the three and four body systems are shown in Figure \ref{fig:4_Body_C3}.

\begin{figure}[H]
    \centering
    \subfigure[]{\includegraphics[width=8.6 cm]{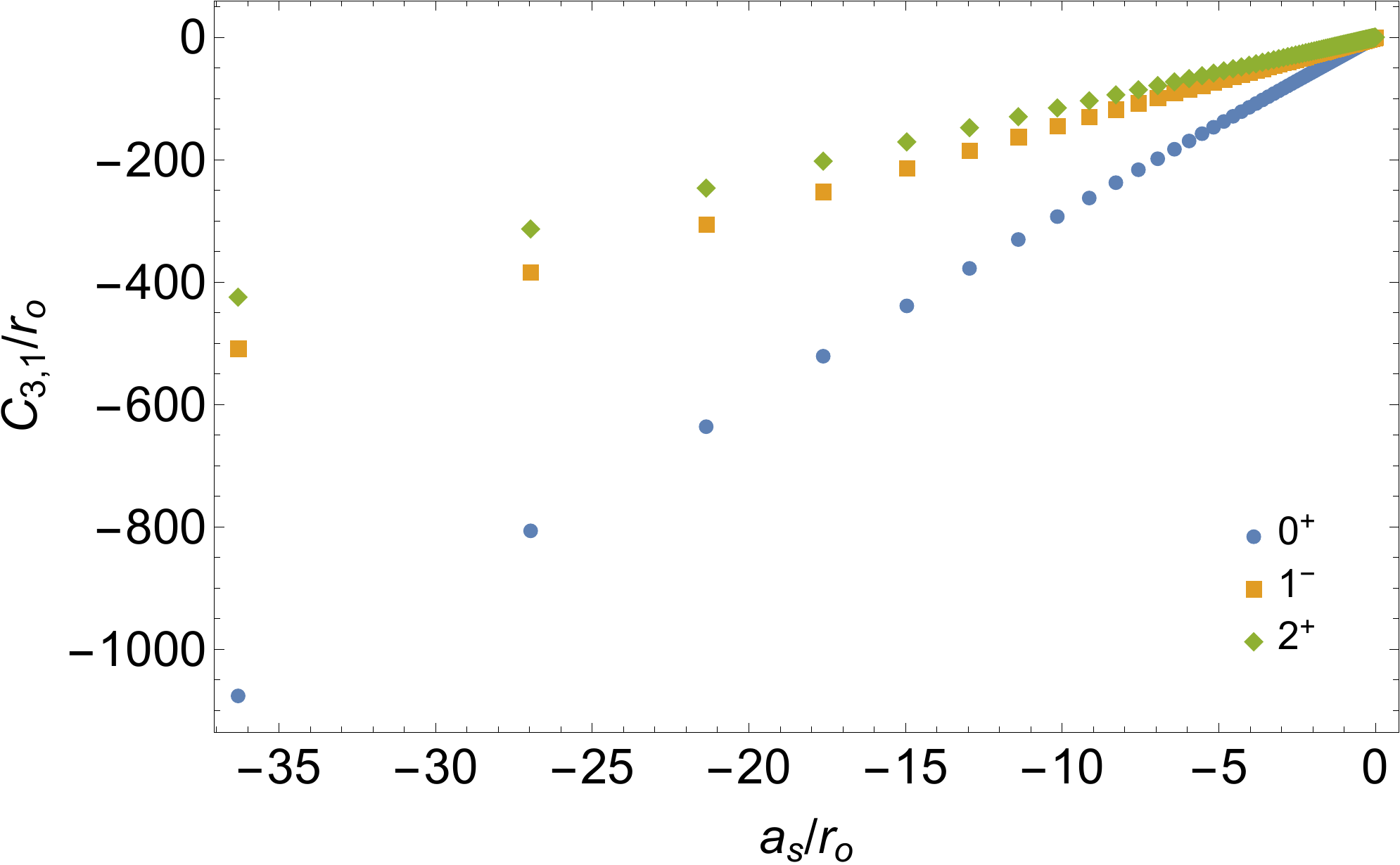}\label{fig:3_Body_C3_as}}

    \subfigure[]{\includegraphics[width=8.6 cm]{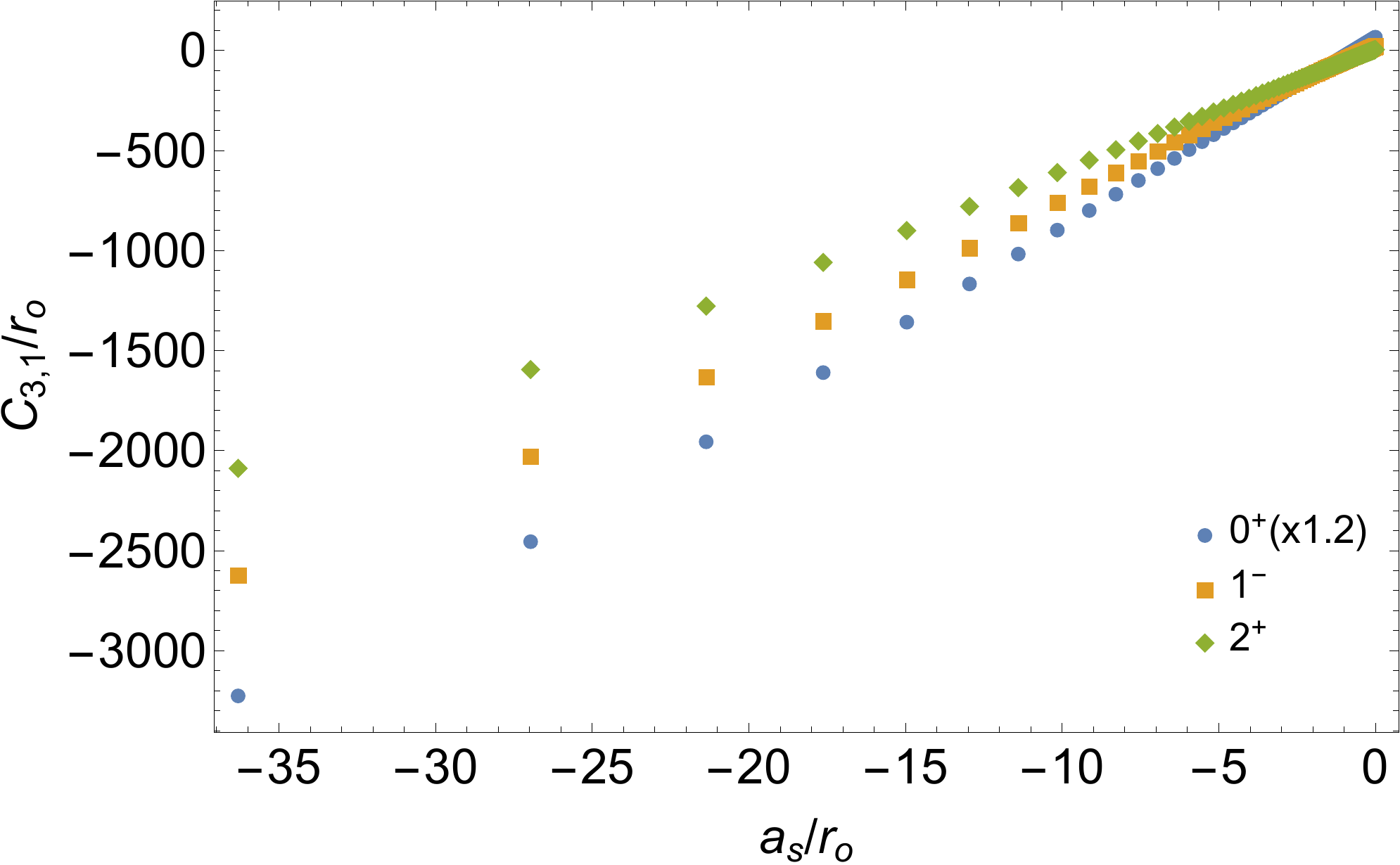}\label{fig:4_Body_C3_as_uudd}}
    
    \subfigure[]{\includegraphics[width=8.6 cm]{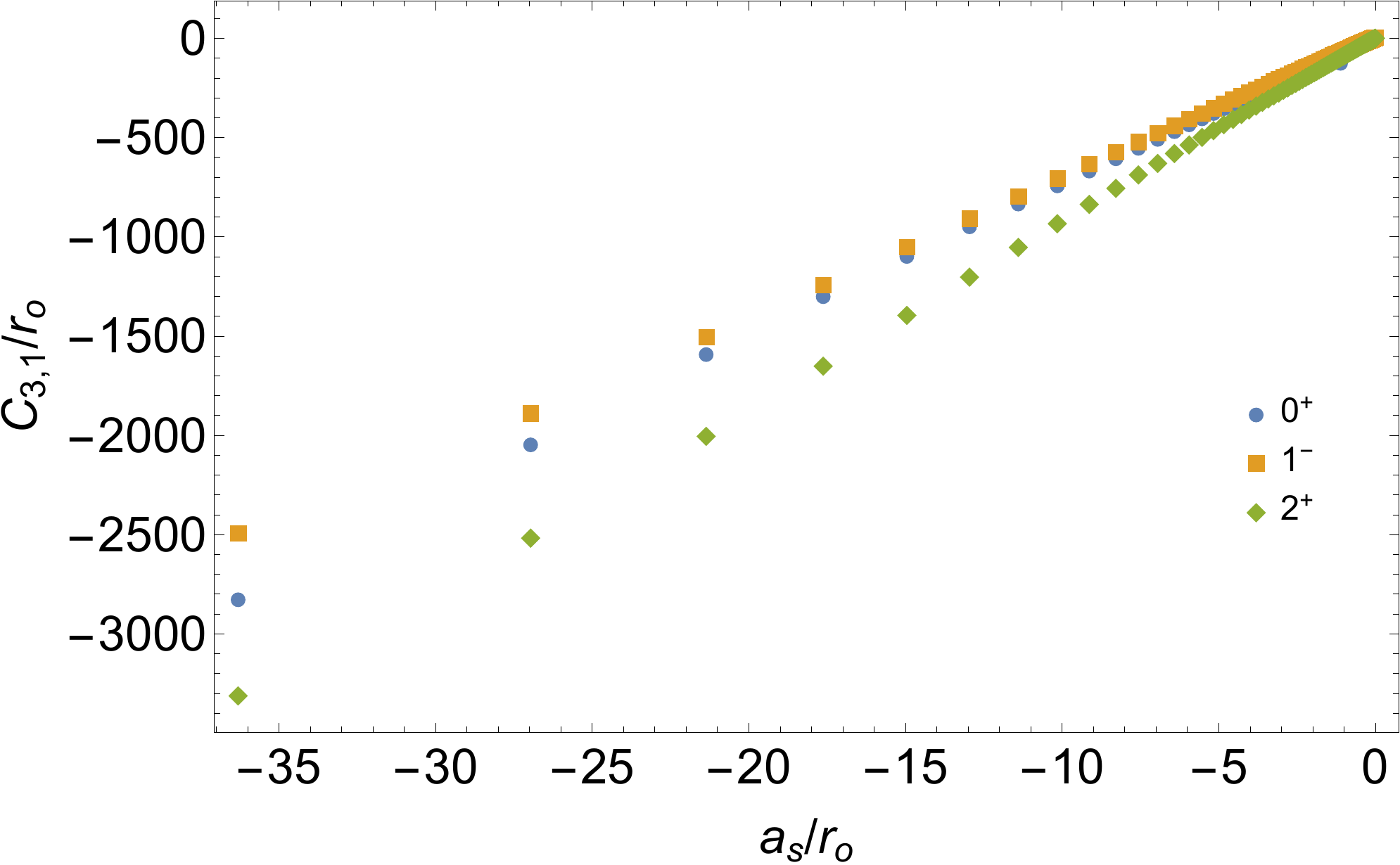}\label{fig:4_Body_C3_as_uuud}}
    
    \caption{The long--range coefficient $C_{3,1}/r_{0}$ versus $a_{s}/r_{0}$ in the lowest Born--Oppenheimer potential for the $(\uparrow\uparrow\downarrow)$ (a), $(\uparrow\uparrow\downarrow\downarrow)$ (b), and $(\uparrow\uparrow\uparrow\downarrow)$ (c) equal mass systems are shown for the symmetries $L^{\pi}$=$0^{+}$ (circles), $1^{-}$ (squares), and $2^{+}$ (diamonds). For values of the $s$--wave scattering length near the unitary limit ($|a_s/r_0|>10$), $C_{3,\nu}$ grows in proportion to $a_s$, in accordance with Eq. \eqref{eq:longrangeform}. These results are for a single Gaussian two--body interaction of width $r_0$. The data for the other hyperspherical channels yield qualitatively similar results to the lowest channel in each symmetry, thus are not shown here.}
\end{figure}

In Fig. \ref{fig:4_Body_C3}, there are numerous curves plotted as a function of $R/r_0$ that go to a constant at $R/r_0\rightarrow\infty$. Each curve, going from top to bottom, represents a different $s$--wave scattering length going from $|a_s|=0$ to $|a_s|\rightarrow\infty$. The curve $C_{3,\nu}(R)$ for $|a_s|\rightarrow\infty$ is linear due to the shift in $l_{\mathrm{eff}}$ at the unitary limit. A similar behavior in the function $C_{3,\nu}(R)$ is observed in the other hyperradial channels and symmetries not shown in Fig. \ref{fig:4_Body_C3}. The universal constant for each system and symmetry is determined through performing fitting procedures to the curves shown in Figs. \ref{fig:3_Body_rsqU_as} and \ref{fig:4_Body_C3}. From this curve fitting procedure, where each curve is fit to an inverse power--law of varying order, the universal constant $\mathcal{C}_{\nu}$ for the lowest Born--Oppenheimer potential, along with the reduced angular momentum quantum number $l_u$, is extracted. In Figures \ref{fig:3_Body_C3_as}--\ref{fig:4_Body_C3_as_uuud}, plots of $C_{3,1}/r_0$ versus $a_s/r_0$ for a single Gaussian two--body interaction are shown for the three different $N$--body systems with panels $a$--$c$ representing the $\uparrow\uparrow\downarrow$, $\uparrow\uparrow\downarrow\downarrow$, and $\uparrow\uparrow\uparrow\downarrow$ spin configurations respectively. In each panel, results are given for the three natural parity symmetries $0^+$ as circles, $1^-$ as squares, and $2^+$ as diamonds. The behavior of the dependence of $C_{3,1}$ on the scattering length is linear for large values of $|a_{s}/r_{0}|$, thus through fitting a linear function to the data shown in Figs. \ref{fig:3_Body_C3_as}--\ref{fig:4_Body_C3_as_uuud} for the lowest hyperradial channel and for the higher hyperradial channels (not shown), the universal parameters $\mathcal{C}_{\nu}$ are extracted from the slope of the fit and given in Table \ref{table:C3_Coeffs_Table}.

\begin{table}[H]
\caption{Long--range coefficients for the three-- and four--body equal--mass systems in the lowest few hyperradial channels for different symmetries. The columns specify the symmetry $L^{\pi}$, the hyperspherical channel index $\nu$ corresponding to a channel that gives a reduced value of $l_{\mathrm{eff}}$ at unitarity, the non--interacting ($l_{\mathrm{eff}}$) and unitary ($l_{u}$) angular momentum quantum numbers, and the universal $1/R^3$ coefficient $\mathcal{C}_{\nu}$. The label $l_{u}^{(\mathrm{ref.})}$ indicates the value of $l_{u}$ extracted from various references. The error bars are estimated based on convergence of the basis used and on curve fitting at large $R$.}
\label{table:C3_Coeffs_Table}
\begin{ruledtabular}
\begin{tabular}{cccccc}
    \multicolumn{6}{c}{\bf{$(\uparrow\uparrow\downarrow)$ system}}\\
    \hline
    $L^{\pi}$ & $\nu$ & $l_{\mathrm{eff}}$ & $l_u$ & $l_{u}^{\mathrm{ref.}}$ & $C_{\nu}$\\
    \hline
    $0^+$ & $1$ & $7/2$ & $1.668(2)$ & $1.6662^{\text{\cite{PhysRevLett.97.150401}}}$, $1.682^{\text{\cite{vonStecher2007prl}}}$  & $29.47(5)$\\
    $~$ & $2$ & $11/2$ & $4.628(2)$ & $4.6274^{\text{\cite{PhysRevLett.97.150401}}}$ & $43.37(5)$\\
    $~$ & $3$ & $15/2$ & $6.616(2)$ & $6.6145^{\text{\cite{PhysRevLett.97.150401}}}$ & $62.3(5)$\\
    $~$ & $5$ & $19/2$ & $8.338(5)$ & $8.3323^{\text{\cite{PhysRevLett.97.150401}}}$ & $147(5)$\\
    \hline
    $1^-$ & $1$ & $5/2$ & $1.273(2)$ & $1.2727^{\text{\cite{PhysRevLett.97.150401}}}$, $1.275^{\text{\cite{vonStecher2007prl}}}$ & $14.19(5)$\\
    $~$ & $2$ & $9/2$ & $3.859(2)$ & $3.85825^{\text{\cite{PhysRevLett.97.150401}}}$ & $14.63(5)$\\
    $~$ & $4$ & $13/2$ & $5.217(2)$ & $5.21643^{\text{\cite{PhysRevLett.97.150401}}}$ & $74.91(5)$\\
    $~$ & $7$ & $17/2$ & $7.556(5)$ & $7.553^{\text{\cite{PhysRevLett.97.150401}}}$ & $98.4(5)$\\
    \hline
    $2^+$ & $1$ & $7/2$ & $2.605(2)$ & $2.60498^{\text{\cite{PhysRevLett.97.150401}}}$ & $11.56(5)$\\
    $~$ & $2$ & $11/2$ & $4.296(2)$ & $4.29541^{\text{\cite{PhysRevLett.97.150401}}}$ & $54.68(5)$\\
    $~$ & $5$ & $15/2$ & $6.734(2)$ & $6.73883^{\text{\cite{PhysRevLett.97.150401}}}$ & $40.52(5)$\\
    $~$ & $9$ & $19/2$ & $8.340(5)$ & $8.3371^{\text{\cite{PhysRevLett.97.150401}}}$ & $136.5(5)$\\
    \hline
    \multicolumn{6}{c}{\bf{$(\uparrow\uparrow\downarrow\downarrow)$ system}}\\
    \hline
    $L^{\pi}$ & $\nu$ & $l_{\mathrm{eff}}$ & $l_u$ & $l_{u}^{\mathrm{ref.}}$ & $C_{\nu}$\\
    \hline
    $0^+$ & $1$ & $5$ & $2.02(2)$ & $2.028^{\text{\cite{YinBlume2015pra}}}$ & $72.0(3)$\\
    $~$ & $2$ & $7$ & $4.45(5)$ & $4.441^{\text{\cite{PhysRevA.85.033634}}}$ & $138(2)$\\
    $~$ & $3$ & $7$ & $5.03(2)$ & $5.029^{\text{\cite{PhysRevA.85.033634}}}$ & $77.0(5)$\\
    $~$ & $4$ & $7$ & $5.36(2)$ & $5.348^{\text{\cite{PhysRevA.85.033634}}}$ & $59.6(5)$\\
    \hline
    $1^-$ & $1$ & $6$ & $4.11(2)$ & $4.0978^{\text{\cite{PhysRevA.85.033634}}}$ & $73.7(3)$\\
    $~$ & $2$ & $6$ & $4.18(2)$ & $4.1758^{\text{\cite{PhysRevA.85.033634}}}$ & $46.1(3)$\\
    $~$ & $3$ & $6$ & $4.75(2)$ & $4.7305^{\text{\cite{PhysRevA.85.033634}}}$ & $45(2)$\\
    $~$ & $4$ & $8$ & $5.73(5)$ & $5.669^{\text{\cite{PhysRevA.85.033634}}}$ & $141(2)$\\
    \hline
    $2^+$ & $1$ & $5$ & $2.95(2)$ & $2.9185^{\text{\cite{PhysRevA.85.033634}}}$ & $59.0(3)$\\
    $~$ & $2$ & $7$ & $4.54(2)$ & $4.539^{\text{\cite{PhysRevA.85.033634}}}$ & $114(2)$\\
    $~$ & $3$ & $7$ & $5.04(2)$ & $5.039^{\text{\cite{PhysRevA.85.033634}}}$ & $96.3(5)$\\
    $~$ & $4$ & $7$ & $5.64(2)$ & $5.6288^{\text{\cite{PhysRevA.85.033634}}}$ & $64.1(5)$\\
    \hline
    \multicolumn{6}{c}{\bf{$(\uparrow\uparrow\uparrow\downarrow)$ system}}\\
    \hline
    $L^{\pi}$ & $\nu$ & $l_{\mathrm{eff}}$ & $l_u$ & $l_{u}^{\mathrm{ref.}}$ & $C_{\nu}$\\
    \hline
     $0^+$ & $1$ & $7$ & $5.35(2)$ & $5.3466(1)^{\text{\cite{stecher2009PRA}}}$ & $75.8(3)$\\
     $~$ & $2$ & $9$ & $6.88(2)$ & $6.8637^{\text{\cite{PhysRevA.85.033634}}}$ & $134(2)$\\
     $~$ & $3$ & $9$ & $7.85(2)$ & $7.8409^{\text{\cite{PhysRevA.85.033634}}}$ & $98(1)$\\
     $~$ & $4$ & $9$ & $8.35(2)$ & $8.3484^{\text{\cite{PhysRevA.85.033634}}}$ & $36(2)$\\
    \hline
     $1^-$ & $1$ & $6$ & $4.17(2)$ & $4.1770^{\text{\cite{PhysRevA.85.033634}}}$ & $69.3(3)$\\
     $~$ & $2$ & $6$ & $4.73(1)$ & $4.7300^{\text{\cite{PhysRevA.85.033634}}}$ & $45.2(5)$\\
     $~$ & $3$ & $8$ & $5.81(2)$ & $5.8068^{\text{\cite{PhysRevA.85.033634}}}$ & $125(3)$\\
     $~$ & $4$ & $8$ & $6.75(3)$ & $6.7219^{\text{\cite{PhysRevA.85.033634}}}$ & $92.2(5)$\\
    \hline
     $2^+$ & $1$ & $7$ & $5.03(2)$ & $5.0385^{\text{\cite{PhysRevA.85.033634}}}$ & $92.2(5)$\\
     $~$ & $2$ & $7$ & $5.74(2)$ & $5.7208^{\text{\cite{PhysRevA.85.033634}}}$ & $60.8(5)$\\
     $~$ & $3$ & $7$ & $5.94(2)$ & $5.9242^{\text{\cite{PhysRevA.85.033634}}}$ & $40.2(5)$\\
     $~$ & $5$ & $9$ & $7.31(2)$ & $7.2742^{\text{\cite{PhysRevA.85.033634}}}$ & $131(3)$\\
\end{tabular}
\end{ruledtabular}
\end{table}

\begin{figure}[H]
    \centering
    \includegraphics[width=8.6 cm]{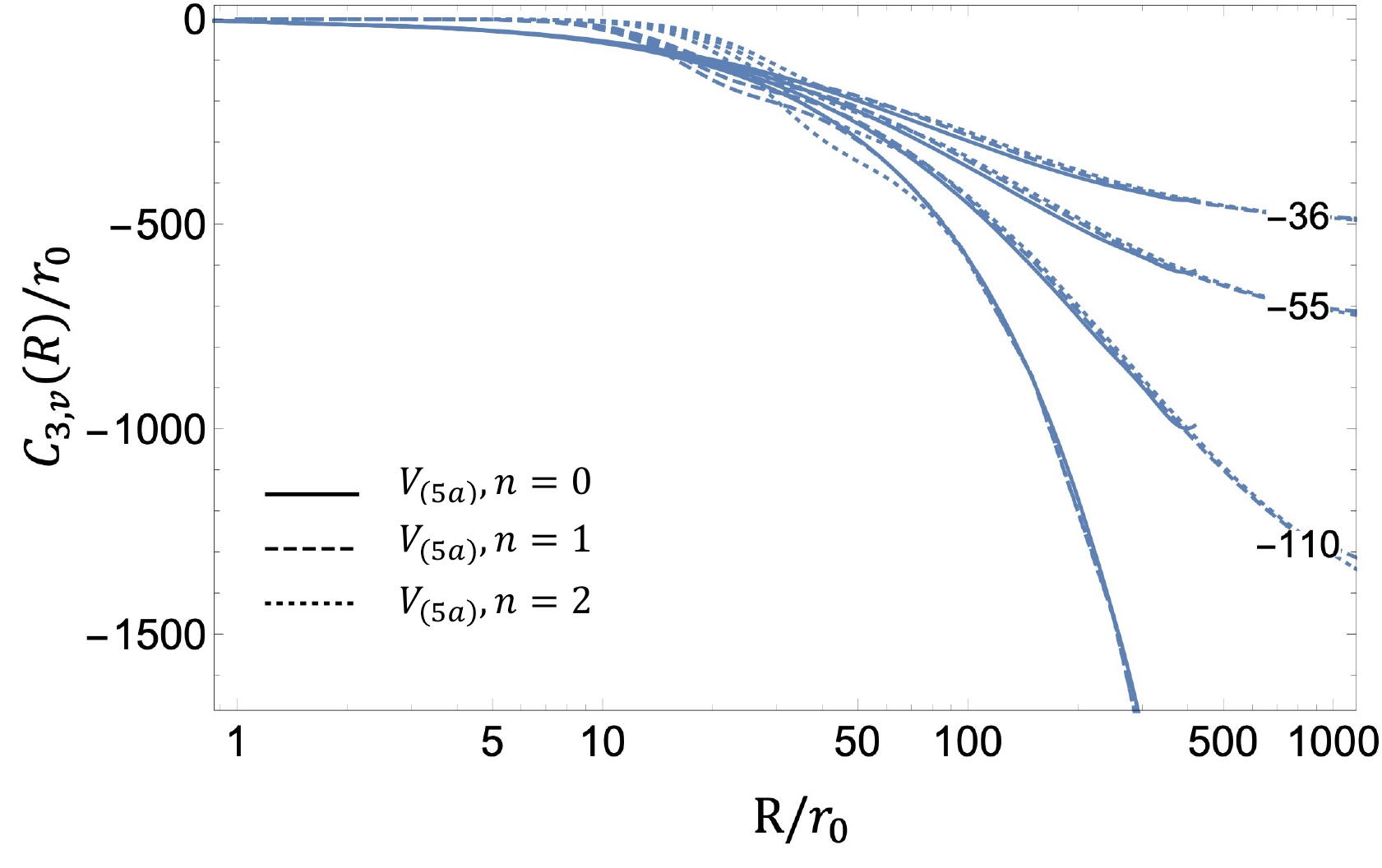}
    
    \caption{\label{fig:C3_comp} A comparison of the long--range coefficient $C_{3,1}$ versus $a_{s}$ in the lowest Born--Oppenheimer potential for the $(\uparrow\uparrow\downarrow)$ equal mass three--body system is shown for three different two--body interactions given in Eq. \eqref{eq:interactions}. The parameters for the interactions are given in Table \ref{table:interaction_params} for $s$--waves. The coefficient $C_{3,1}$ and the hyperradius $R$ are rescaled by the range of the single Gaussian interaction $r_0$. The set of three curves are labeled by the ratio of $a_s/r_0$ shown on the right hand side. Universality kicks--in for large hyperradius, is indicated by the different interactions converging to one curve for each scattering length.}
\end{figure}

Universality in the hyperradial potentials at long--range appears for continuum states as a reduction in the angular momentum barrier at infinite scattering length. This is further supported in Figure \ref{fig:C3_comp}, which shows a plot of $C_{3}(R)$ for the lowest hyperradial potential using different two--body interactions. The two body interactions used in Fig. \ref{fig:C3_comp} are of the form given by Eq. \eqref{eq:interactions2}. At large hyperradius, the hyperradial potentials collapse onto one curve, demonstrating long--range universality in these continuum states. Only at small hyperradii, at distances less than $R/r_0\sim20$, does the short range nature of the two--body interaction become important and non--universal behavior emerges. The analysis of the long--range behavior of the hyperradial potentials thus far has been for the individual spin components of the interacting fermions.

In other systems, such as the few--nucleon systems in nuclear physics, the good quantum numbers are either the total orbital angular momentum and spin $L^{\pi}$ and $S$, or the total angular momentum $J^{\pi}$ in an $LS$--coupling scheme. For the three--body system, the symmetries treated are $(L^{\pi},S)=(0^+,1/2)$ and $(1^-,1/2)$. Likewise, for the four--body system, the symmetries treated are $(L^{\pi},S)=(0^+,0)$ and $(1^-,0)$. From the nuclear studies, near unitary physics was explored for both three--body and four--body systems in the symmetries $(L^{\pi},S)=(1^-,1/2)$ and $(0^+,0)$ respectively. The long range universal behavior addressed above manifests itself in few neutron systems as a result of the large neutron--neutron singlet two--body $s$--wave scattering length. This behavior has been studied for the three and four interacting neutron systems \cite{PhysRevLett.125.052501,PhysRevC.103.024004}. Through spin re-coupling, the long--range $1/R^3$ coefficients for a given total spin $S$ in the lowest hyperradial channel can be determined from the values given in Table \ref{table:C3_Coeffs_Table}. Some results for the total spin states are given in Table \ref{table:totalspin} for three-- and four--body systems. The long--range coefficient for given total spin $S$ is typically not relevant for cold atom systems in the presence of a magnetic field since $S$ is not a good quantum number in this case.

In nuclear physics, there is a classification of spin-1/2 particles interacting in the unitarity regime denoted as ``unparticles" that can be described from conformal field theory \cite{PhysRevLett.98.221601,Unnucleus_PNAS}. An example of such systems are clusters of neutrons. The energy dependence of the differential cross--section for an $N$--body system is given as $d\sigma/dE\propto E^{\Delta-5/2}$, where $\Delta$ is the conformal dimension for the $N$--body system. Rewriting the energy dependence in terms of the wavenumber $k$, one can see that from the Wigner threshold law where $d\sigma/dE\propto k^{2l_{\mathrm{eff}}+1}$, $\Delta$ and $l_{\mathrm{eff}}$ are related through the expression $\Delta=l_{\mathrm{eff}}+3$. Using the values of $l_{\mathrm{u}}$ given in Table \ref{table:totalspin}, the conformal dimensions given in \cite{Unnucleus_PNAS} are reproduced for the $N=3,~4$ cases.

\begin{table}[!ht]
\caption{Long--range coefficients for the three--body system with total angular momentum and parity $L^{\pi}$, and total spin $S$. For each $L$ and $S$, the non--interacting angular momentum quantum number $l_{\mathrm{eff.}}$ is given along with the reduced value $l_{u}$ at the $s$--wave unitary limit. The last column gives the numerical results for the scattering length dependent coefficient $\mathcal{C}_{\nu}$ from Eq. \eqref{eq:longrangeform} or the hyperspherical channel $\nu$ that gives a reduced $l_{\mathrm{eff}}$ at unitarity. The error bars are estimated based on convergence of the basis used and on curve fitting at large $R$.}
\centering
\label{table:totalspin}
\begin{ruledtabular}
\begin{tabular}{lllllll}
    $N$ & $(L^{\pi},S)$ & $\nu$ & $l_{\mathrm{eff}}$ & $l_u$ & $l_{u}^{(\mathrm{ref.})}$ & $\mathcal{C}_{\nu}$\\
    \hline
    $3$ & $(1^{-},1/2)$ & $1$ & $5/2$ & $1.275(3)^{\ref{fn:HigginsPRC}}$ & $1.2727^{\text{\cite{PhysRevLett.97.150401}}}$ & $15.1(3)\footnote{Value extracted from \cite{PhysRevC.103.024004}. The $\mathcal{C}_{\nu}$ coefficients are multiplied by $\sqrt{\mu/\mu^{\prime}}$ where $\mu^{\prime}$ is the neutron--neutron reduced mass. \label{fn:HigginsPRC}}$\\
    $~$ & $~$ & $2$ & $9/2$ & $3.861(3)^{\ref{fn:HigginsPRC}}$ & $3.8582^{\text{\cite{PhysRevLett.97.150401}}}$ & $15.2(3)^{\ref{fn:HigginsPRC}}$\\
    $~$ & $~$ & $3$ & $13/2$ & $5.219(3)^{\ref{fn:HigginsPRC}}$ & $5.2164^{\text{\cite{PhysRevLett.97.150401}}}$ & $77.7(3)^{\ref{fn:HigginsPRC}}$\\
    $~$ & $~$ & $4$ & $17/2$ & $17/2$ & $17/2$ & $---$\\
    $~$ & $~$ & $5$ & $17/2$ & $7.555(3)^{\ref{fn:HigginsPRC}}$ & $7.553^{\text{\cite{PhysRevLett.97.150401}}}$ & $108(3)^{\ref{fn:HigginsPRC}}$\\\hline
    $3$ & $(0^{+},1/2)$ & $1$ & $7/2$ & $1.668(5)$ & $1.6662^{\text{\cite{PhysRevLett.97.150401}}}$ & $29.8(5)$\\
    $~$ & $~$ & $2$ & $11/2$ & $4.630(5)$ & $4.6274^{\text{\cite{PhysRevLett.97.150401}}}$ & $44.6(5)$\\
    $~$ & $~$ & $3$ & $15/2$ & $6.63(2)$ & $6.6145^{\text{\cite{PhysRevLett.97.150401}}}$ & $63.0(5)$\\
    $~$ & $~$ & $4$ & $19/2$ & $8.35(2)$ & $8.3323^{\text{\cite{PhysRevLett.97.150401}}}$ & $156(5)$\\\hline
    $4$ & $(0^{+},0)$ & $1$ & $5$ & $2.02(2)^{\ref{fn:HigginsPRC}}$ & $2.0094(1)^{\text{\cite{YinBlume2015pra}}}$ & $86.7(3)^{\ref{fn:HigginsPRC}}$\\
    $~$ & $~$ & $2$ & $7$ & $4.45(2)^{\ref{fn:HigginsPRC}}$ & $4.444(3)^{\text{\cite{stecher2009PRA}}}$ & $156(3)^{\ref{fn:HigginsPRC}}$\\
    $~$ & $~$ & $3$ & $7$ & $5.07(5)^{\ref{fn:HigginsPRC}}$ & $5.029(3)^{\text{\cite{stecher2009PRA}}}$ & $61.1(3)^{\ref{fn:HigginsPRC}}$\\
    $~$ & $~$ & $4$ & $9$ & $6.97(5)^{\ref{fn:HigginsPRC}}$ & $6.863(3)^{\text{\cite{stecher2009PRA}}}$ & $209(3)^{\ref{fn:HigginsPRC}}$\\
    $~$ & $~$ & $5$ & $9$ & $7.26(5)^{\ref{fn:HigginsPRC}}$ & $7.121(3)^{\text{\cite{stecher2009PRA}}}$ & $87.8(3)^{\ref{fn:HigginsPRC}}$\\\hline
    $4$ & $(1^{-},0)$ & $1$ & $6$ & $4.11(2)$ & $4.0978^{\text{\cite{PhysRevA.85.033634}}}$ & $73.6(5)$\\
    $~$ & $~$ & $2$ & $8$ & $5.74(5)$ & $5.667^{\text{\cite{PhysRevA.85.033634}}}$ & $116(5)$\\
    $~$ & $~$ & $3$ & $8$ & $6.57(5)$ & $6.505^{\text{\cite{PhysRevA.85.033634}}}$ & $81.2(5)$\\
\end{tabular}
\end{ruledtabular}
\end{table}

\subsection{Low--energy behavior of elastic phase shifts}
The long--range behavior of the Born--Oppenheimer potentials discussed in the previous section leads to important implications at low collision energies. As described, the long--range behavior of the adiabatic potentials for $s$--wave two--body interactions near the unitary limit, specifically near the first $s$--wave pole in $a_s$, the adiabatic potentials exhibit a $1/R^3$ long--range tail proportional to the scattering length $a_s$, represented by Eq. \eqref{eq:longrangeform}. Using the Born approximation, the low--energy elastic $N$-body scattering phase shift for a given angular momentum $l_{\mathrm{eff}}$ is
\begin{equation}
    \label{eq:bornapprox}
    \delta_{l_{\mathrm{eff}}}(k)=-[C/(2l_{\mathrm{eff}}(l_{\mathrm{eff}}+1))]k
\end{equation}
where $C$ is the coefficient of the $1/R^3$ term in Eq. \eqref{eq:longrangeform}. Moreover, as was mentioned above, the squared $S$-matrix element for recombination is proportional to $k^{2l_{\mathrm{eff}}+1}$ in the generalized Wigner threshold law for that process \cite{RAU_1984,Sadeghpour_2000}, when starting from an initial channel with centrifugal potential barrier coefficient $l_{\mathrm{eff}}(l_{\mathrm{eff}}+1)$.

\begin{figure}[H]
    \centering
    \includegraphics[width=8.6 cm]{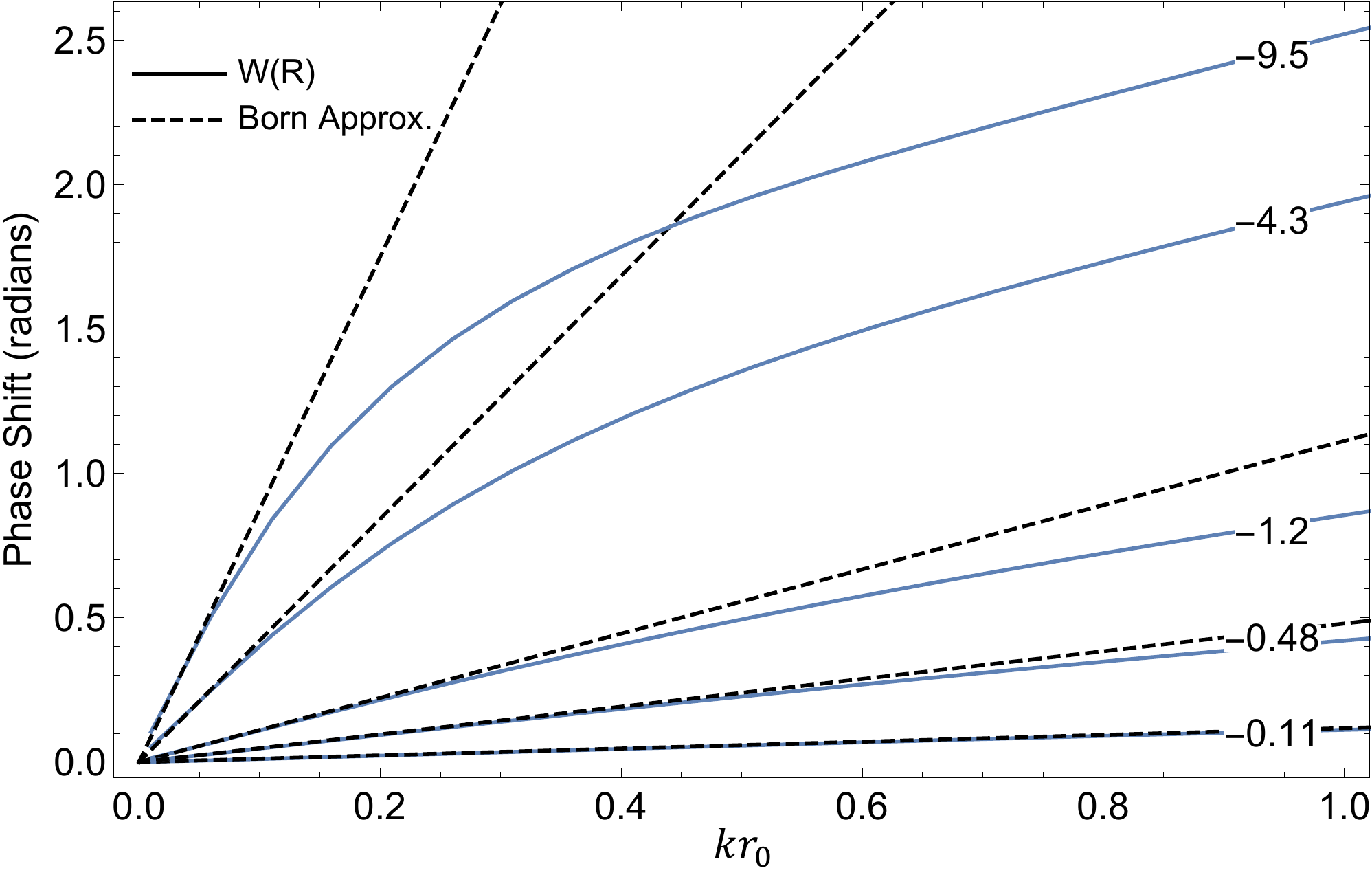}
    
    \caption{\label{fig:phaseshift_sample} Single--channel elastic phase shifts for collisions in the lowest hyperspherical continuum channel. Each curve is computed using a different two--body $s$--wave scattering length, where the ratio $a_s/r_0$ is given for a Gaussian interaction and displayed on the right. The low energy behavior is governed by the characteristics of the long--range form of the hyperradial potential, see Eq. \eqref{eq:longrangeform}. The $1/R^3$ term in the long--range form gives rise to a linear dependence on wave number $k$ of the phase shift, in accordance with the Wigner threshold law. This linear dependence, derived using the born approximation in Eq. \eqref{eq:bornapprox}, is shown as the black dashed lines.}
\end{figure}

Figure \ref{fig:phaseshift_sample} shows a sampling of elastic phase shifts for a Gaussian two--body interaction for different scattering lengths. The four--body elastic phase shift is given as a function of the wave number $k$ for different negative s--wave scattering lengths to highlight the changes in the low--energy behavior as the scattering length increases from zero to the unitary limit. Also shown in Fig. \ref{fig:phaseshift_sample} is the low--energy limit derived from the first Born approximation, represented by the black dashed lines. The implications of this linear behavior of the phase shift with wave number comes into the density of final continuum states, defined through the energy derivative of the phase shift for a single--channel calculation or through the time delay matrix in a multi--channel calculation. From Eq. \eqref{eq:bornapprox}, the low energy density of states, related through $d/dE$, results in an energy enhancement near threshold of $1/E$.

\section{Interacting fermions at the p--wave unitary limit\label{sec:section3}}
\begin{figure}[H] 
\hspace{-0.0in}\includegraphics[width=8.6 cm] {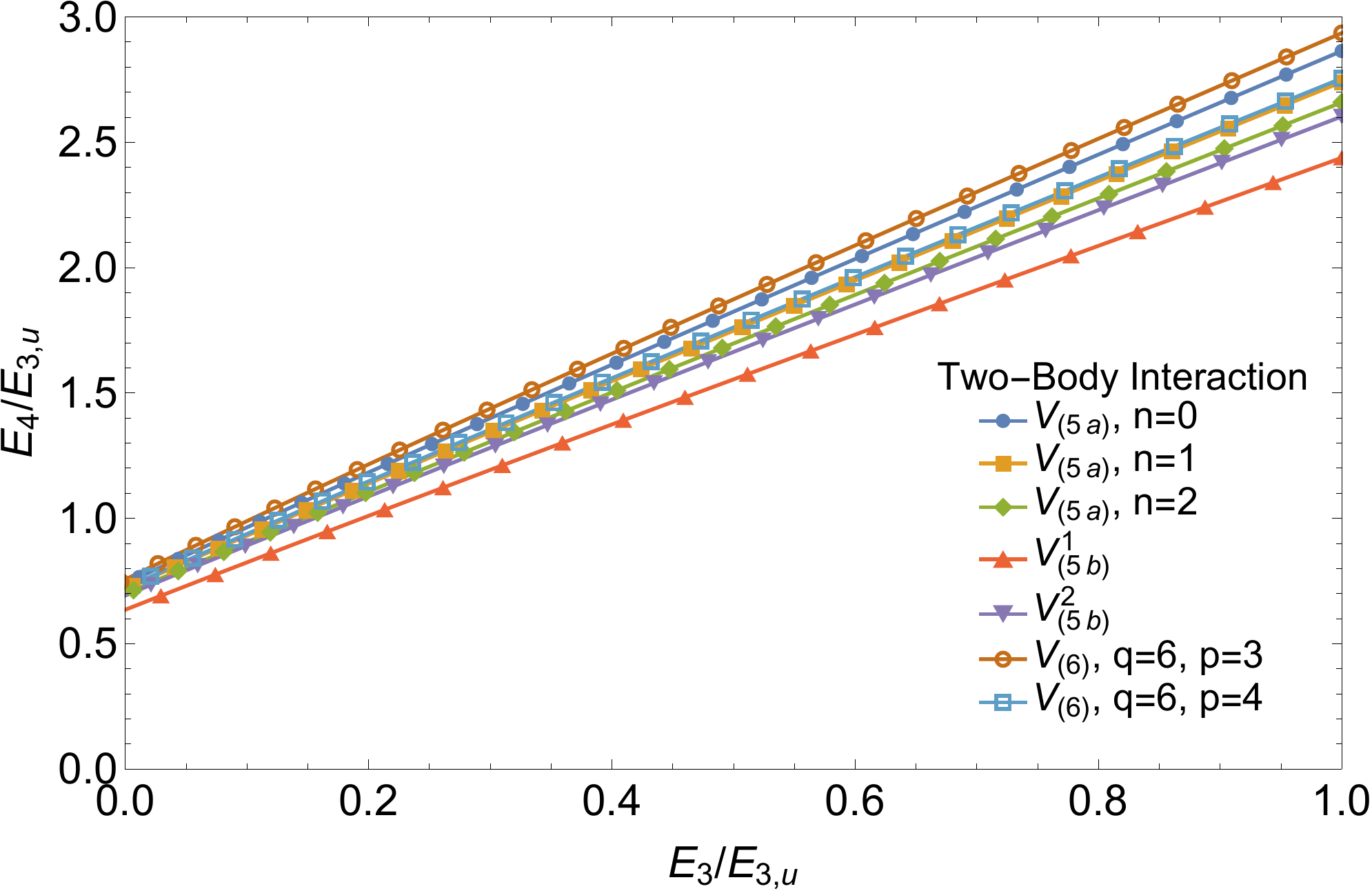}
%\vskip-3.5in 
\caption{\label{fig:Tjon_Plot} A Tjon plot of the correlation between the universal three--body binding energy in the $L^{\pi}=1^-$ and the lowest four--body binding energy in the $L^{\pi}=0^+$ symmetry. Each point type corresponds to a different two--body interaction, given in the legend and labeled by the equation number. The parameters for the interactions are given in Table \ref{table:interaction_params} for $p$--waves. The axes are rescaled by the trimer energy at the unitary limit to place all of the data sets on the same scale. }
\end{figure}

The second type of few--body fermionic systems looked at in this study are the interactions of identical spin--polarized fermions, which interact through $p$--wave interactions. The focus of this study is to understand four--body properties for these systems and relate them to the corresponding three--body properties, which have been studied in past theoretical and experimental works \cite{PhysRevA.105.013308,Suno_2003,PhysRevLett.120.133401,PhysRevA.98.020702,PhysRevLett.98.200403,PhysRevLett.100.143201,Ketterle_2021}. In spin--polarized fermionic systems at unitarity, it has been shown in a number of studies that there is no Efimov effect that emerges at unitarity unlike for bosonic systems \cite{PhysRevA.86.012711,PhysRevA.105.013308,PhysRevA.92.033626,PhysRevLett.109.230404}. Short--range effects on three--and four--body properties are investigated for the spin--polarized configuration in the $1^{-}$ and $0^{+}$ symmetries, respectively. Through solving the full Schr\"odinger equation over all space, the three--body binding energy is obtained as a function of the $p$--wave scattering volume for different two--body interactions. Likewise, the lowest four--body bound state energy has been obtained in the $L^{\pi}$=$0^+$ symmetry as a function of the $p$--wave scattering volume. Correlations between the three--body bound state and the four--body state at the $p$--wave unitary limit are made for different two--body interactions. The results are presented in a Tjon plot, similar to ones shown for bosonic systems in past publications \cite{PhysRevA.99.013613,PhysRevA.74.063604,Blume2015EfimovPhys,PhysRevA.92.033626,PhysRevLett.109.230404}, and are shown in Figure \ref{fig:Tjon_Plot}.

Figure \ref{fig:Tjon_Plot} is a Tijon plot that shows the correlation between the three--body trimer energy in the $1^{-}$ symmetry and the four--body tetramer energy in the $0^{+}$ for spin--polarized fermions in each case (i.e the total spin is $S=3/2$ and $S=2$ for the three and four body systems respectively). From Fig. \ref{fig:Tjon_Plot}, the three-- and four--body ground--state energies are given for different two--body interactions, provided in Eqs. \eqref{eq:interactions} and \eqref{eq:vanderwaals}. The respective three--body and four--body energies are rescaled by the trimer value at the $p$--wave unitary limit in order to display the results from each interaction on the same scale. The correlation between the trimer and tetramer energies are linear for each interaction type. A linear fit was performed over the average of the two--body interaction models to determine an effective correlation between the trimer and tetramer states. The linear fit parameters are given in the figure and the correlation has been determined (with an $R$--squared value of 0.9754) to be $E_4^{0^+}=2.024(11)E_3^{1^-}+0.718(6)E_{3,u}^{1^-}$, where $E_4^{0^+}$ is the tetramer energy for $L^{\pi}=0^+$, $E_3^{1^-}$ is the trimer energy for $L^{\pi}=1^-$, and $E_{3,u}^{1^-}$ is the trimer energy at the $p$--wave unitary limit. There is a similar relationship between the trimer and tetramer states in four--boson systems at the $s$--wave unitary limit, in relation to the Efimov effect. In the bosonic systems, there exist two tetramer bound states for every hyperspherical potential describing the two--body fragmentation to an Efimov trimer+free particle \cite{Platter2004,StecherNature,DeltuvaFBS2013}. The relationship between the two universal tetramers and the corresponding Efimov trimer at unitarity was determined to be $E_{4B}^{n,m}=c_{m}E_{3B}^{n}$, where $c_1=4.58$ and $c_2=1.01$, as described in Eq. (2) of \cite{StecherNature}. Using momentum--space transition operators to study these universal tetramers for unitary bosons, Deltuva computed the universal coefficients to be $c_1=4.610(1)$ and $c_2=1.00227(1)$ \cite{DeltuvaFBS2013}. In the case of fermions, the scaling coefficient of the trimer energy is smaller than the scaling factor of the ground state for bosons by almost a factor of 2. This difference can be interpreted as resulting from the Pauli repulsion effects seen in fermionic systems that are absent from bosonic systems.

Unlike the four--boson case, the four--fermion tetramer is not universal, i.e. its energy depends on the short range interaction. Table \ref{table:pwave_properties} shows this ratio for different two--body interactions. The reason for this non--universal behavior is similar to that in the four--boson case for the two tetramers in the lowest hyperradial potentials in \cite{StecherNature}. The hyperradial potentials that support the ground--state tetramers in the fermion case have potential minima on the order of the range of the interaction. Thus, the mean distance between any two fermions in the tetramer is of the order of the two--body interaction range, so it stands to reason that the trimer and tetramer energies will vary depending on the short--range behavior of the interaction, similar to that found in the lowest hyperradial potential for the boson case.

\begin{table}[ht]
\caption{Energy and scattering volume ratios at the $p$--wave unitary limit for the spin--polarized $(\uparrow\uparrow\uparrow\uparrow)$ system in the $0^{+}$ symmetry. Each column represents a different two--body interaction from Eqs. \eqref{eq:interactions} and \eqref{eq:vanderwaals}, where the parameters are given in Table \ref{table:interaction_params} for $p$--waves. Each row labels the ratio between four-- and three--body properties, where row 1 gives the ratio of the $0^{+}$ tetramer energy to the $1^{-}$ trimer energy at unitarity, row 2 gives the ratio of overall scaling factors ($\beta_{p,4}^{0^+}/\beta_{p,3}^{1^-}$) of the two--body interaction (i.e. $V^{\prime}(r)=\beta V(r)$) needed to bind the trimer and tetramer, and row 3 gives the ratio of effective $p$--wave scattering lengths $a_p=V_p^{1/3}$ at which tetramer and trimer states first form at zero energy.  }
\centering
\label{table:pwave_properties}
\begin{ruledtabular}
\begin{tabular}{llllllll}
    $V_{ij}$ & $V_{\eqref{eq:interactions2}}^{n=0}$ & $V_{\eqref{eq:interactions2}}^{n=1}$ & $V_{\eqref{eq:interactions2}}^{n=2}$ & $V_{\eqref{eq:interactions3}}^{(1)}$ &
    $V_{\eqref{eq:interactions3}}^{(2)}$ &
    $V_{\eqref{eq:vanderwaals}}^{p=3}$ & $V_{\eqref{eq:vanderwaals}}^{p=4}$\\
    \hline
    $E_{4,u}^{0^{+}}/E_{3,u}^{1^{-}}$ & $2.864$ & $2.740$ & $2.659$ & $2.603$ & $2.446$ & $2.936$ & $2.755$\\
    $\beta_{p,4}^{0^+}/\beta_{p,3}^{1^{-}}$ & $0.935$ & $0.941$ & $0.944$ & $0.944$ & $0.953$ & $0.933$ & $0.938$\\
    $a_{p,4}^{0^+}/a_{p,3}^{1^{-}}$ & $0.875$ & $0.874$ & $0.876$ & $0.882$ & $0.874$ & $0.891$ & $0.887$\\
\end{tabular}
\end{ruledtabular}
\end{table}

Other quantities of interest are the points at which the trimer and tetramer states transition from being a resonance in the continuum to being a bound state. These transition points are characterized by the ratio of the effective $p$--wave scattering lengths, $a_p=V_p^{1/3}$, at which this transition occurs for the three-- and four--body systems. This ratio is numerically determined here through a full diagonalization of the Hamiltonian for the different two--body interactions used to characterize the energy ratio and is shown in the second row of Table \ref{table:pwave_properties}. From these results, the ratio $a_{p,4}^{0^{+}}/a_{p,3}^{1^{-}}\sim0.88$ is a universal quantity, as it is found to be insensitive to the short--range behavior of the interaction, with a standard deviation of $0.01$, i.e. $1.0\%$ of the mean.

The quantity $a_{p,4}/a_{p,3}$ has importance for four--body recombination processes. In this case, the relevant recombination process is $A+A+A+A\leftrightarrow A_3+A$ for $L^{\pi}=0^{+}$, where $A$ is a fermion. Given the two--body $p$--wave scattering volume $V_{p,3}^{AA}$, with $a_{p,3}=[V_{p,3}^{AA}]^{1/3}$, where the $A_3$ trimer transitions from resonant to bound, there will be an enhancement in the four--body recombination for an ultracold gas into deep dimers or trimers at scattering volumes $V_{p,4}^{AA}$, with $a_{p,4}=[V_{p,4}^{AA}]^{1/3}$, where the tetramer $A_4$ transitions from resonant to bound. The ratio $a_{p,4}^{0^{+}}/a_{p,3}^{1^{-}}$ is universal with small deviations (two standard deviations of the mean, or $\sim2\%$) due to short range effects (see Table \ref{table:pwave_properties}). The location of the enhancement in the four--body recombination process where the four--body state becomes resonant with the continuum is governed by $a_{p,3}$ of the system and should be found in the range $0.87a_{p,3}^{1^{-}}<a_{p,4}^{0^{+}}<0.89a_{p,3}^{1^{-}}$ or in terms of the $p$--wave scattering volume, $0.66V_{p,3}^{1^{-}}<V_{p,4}^{0^{+}}<0.70V_{p,3}^{1^{-}}$. It should be noted that the small deviations resulting from short--range effects are of the same order of magnitude as those found in the $N$--boson case when looking at the lowest Efimov states \cite{von_Stecher_2010,PhysRevLett.107.200402,PhysRevLett.102.140401}.

\begin{figure}[H] 
\hspace{-0.0in}\includegraphics[width=8.6 cm] {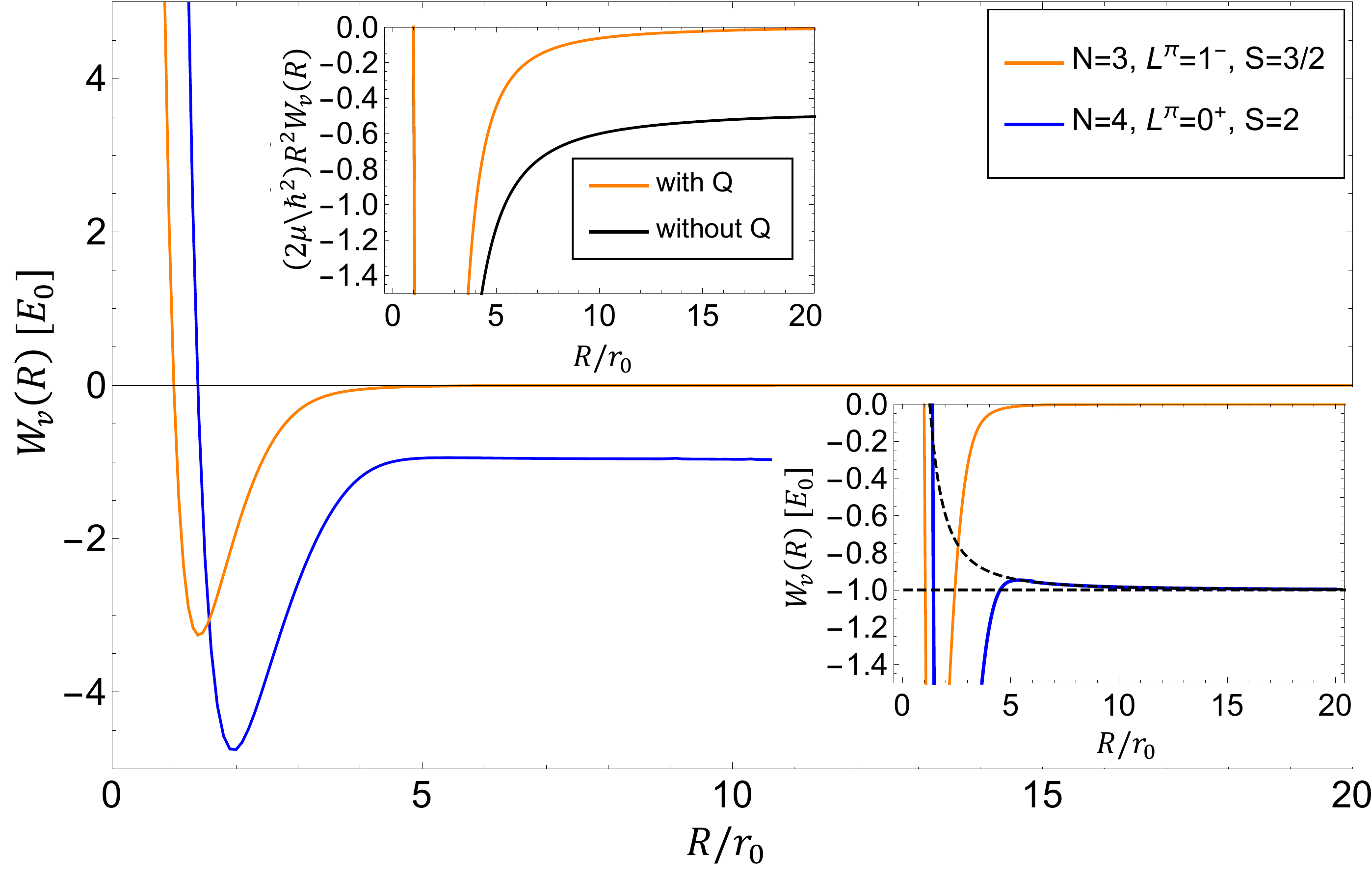}
%\vskip-3.5in 
\caption{\label{fig:3B_4B_pots} Lowest hyperradial potentials for the three-- and four--identical fermionic systems in the spin--polarized state at the $p$--wave unitary limit. The upper--inset plot shows the three--body potential rescaled by $R^2$ with and without the second--derivative non--adiabatic coupling. This highlights the long--range behavior, specifically showing the negative, non--zero coefficient of $R^{-2}$ without the diagonal coupling added, indicating Efimov physics. However, this is not a true Efimov feature, as this coefficient vanishes after including the non--adiabatic coupling.}
\end{figure}

These identical fermionic systems interacting at the $p$--wave unitary limit are further studied by calculating the adiabatic potential curves as functions of the hyperradius. To verify the relation between the trimer energy and tetramer energy given in Table \ref{table:pwave_properties}, the lowest few hyperradial potential curves for the spin--polarized ($S=3/2$ and $S=2$) three-- and four--body systems were computed, along with the diagonal non--adiabatic second--derivative correction to provide an upper and lower bound to these binding energies. Figure \ref{fig:3B_4B_pots} shows the lowest three-- and four--body hyperradial potential with the diagonal non--adiabatic correction. In Fig. \ref{fig:3B_4B_pots}, the lowest effective hyperradial potentials for the three-- and four fermionic spin--polarized systems are shown in orange and blue respectively, up to 20 a.u. The two inset plots highlight some features of these effective potentials at large hyperradius related to fragmentation pathways and the Efimov effect. The inset plot on the lower left shows a zoomed--in view of the main figure which highlights the asymptotic behavior of the lowest $0^+$ four--body potential.  This inset shows the behavior of the lowest four--body potential at large hyperradius, which represents the two--body fragmentation of the four--body system into the universal trimer in the $1^-$ state plus a free particle with a relative angular momentum quantum number of $l=1$. This is further emphasized by the comparison of the effective potential with a purely centrifugal barrier (represented by the dashed line) with $l=1$. The horizontal dashed line shows the trimer energy, which aligns with the position of the trimer energy in the well of the three--body potential in the main figure. 

The second inset shows the lowest three--body hyperradial potential multiplied by $R^2$ with and without the second--derivative non--adiabatic coupling term. As shown, the Born--Oppenheimer potential falls off as $1/R^2$ with a negative power of approximately $-0.5$ (i.e. $U_{0}(R)\rightarrow-\frac{\hbar^2}{4\mu R^2}$). From this attractive long--range behavior, the lowest three--body adiabatic potential behaves like the Efimov potential manifested in the three--boson system at $s$--wave unitarity, that has the form $U_{0}(R)=-\frac{\hbar^2}{2 \mu R^2}(s_{0}^2+1/4)$ with $s_{0}=1.0061$. Comparing the $1/R^2$ coefficient for bosons with the coefficient of the adiabatic potential for three fermions, the value of the Efimov parameter is $s_{0}\approx0.5$. Since the value of $s_0$ is greater than zero, this would indicate an Efimov--like effect in this fermionic system. However, it has been determined that no such Efimov effect actually exists in these systems. The non-adiabatic second--derivative coupling exactly cancels this attractive $1/R^2$ term in the adiabatic potential, indicated by the upper line in the inset that goes to zero at large $R$. This indicates that the lowest effective three--body hyperradial potential does not have an Efimov--like behavior at large hyperradius which would lead to an infinite set of bound trimer states. 

\begin{figure}[H] 
\hspace{-0.0in}\includegraphics[width=8.6 cm] {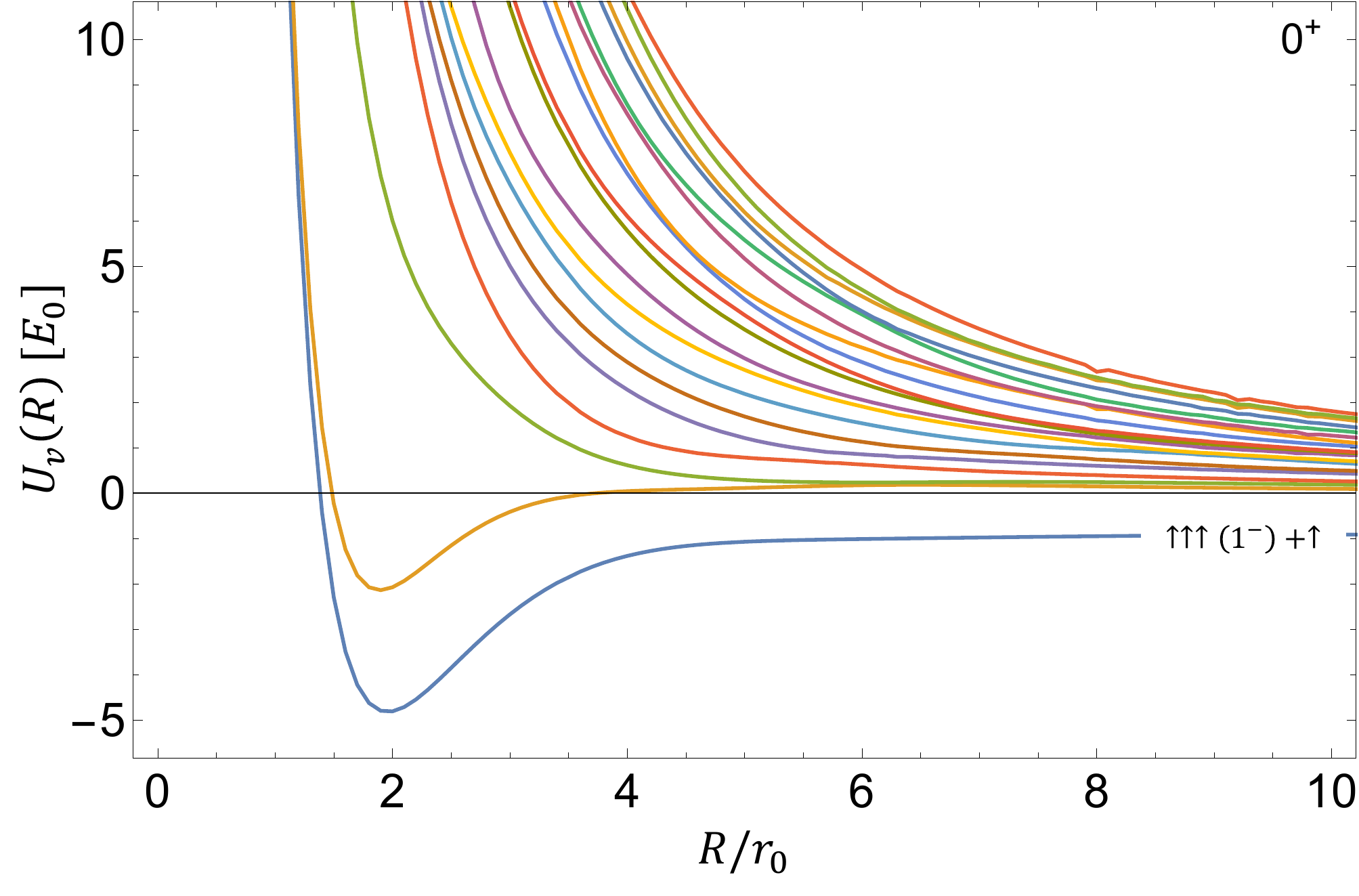}
%\vskip-3.5in 
\caption{\label{fig:4B_uuuu_pu} The hyperradial potentials for the four--fermion system in the $(\uparrow\uparrow\uparrow\uparrow)$ spin configuration. The two--body interaction between $(\uparrow\uparrow)$ spins is set to the first $p$--wave unitary limit. For this set of two--body interactions, there is one two two--body breakup threshold below the four--body continuum, labeled by the fragmentation of the spins and the angular momentum of the trimer.}
\end{figure}

Further evidence for the non--existence of the Efimov effect for three identical spin--polarized fermions is shown in the hyperradial mapping of the four--body case, shown in Figure \ref{fig:4B_uuuu_pu}. The spectrum of hyperradial potentials in the four--body case give further insights into the interactions of identical fermions at unitarity. From Fig. \ref{fig:4B_uuuu_pu}, there is only one fragmentation pathway of the four--body system to a two--body system representing the fragmentation into a $p$--wave universal trimer and a free particle, shown in the lowest potential. If there were a three--body Efimov effect in the identical spin--polarized fermion case, then there should be an infinite number of hyperradial potentials, each one going to one of the Efimov trimers as is the case of bosonic systems \cite{StecherNature}.

There are other interesting features in the spectrum of the four--body system at the $p$--wave unitary limit that should be noted. One of the main features, as described earlier, is the presence of a bound tetramer state in the lowest hyperspherical potential. In the $L^{\pi}=0^+$ symmetry, there is a single tetramer bound state in the lowest potential which asymptotically, at large hyperradius, goes to the trimer plus free particle energy threshold. The binding energy of the spin--polarized tetramer is related to the corresponding trimer energy through a relation described  by a Tjon plot like the one shown in Fig. \ref{fig:Tjon_Plot}. There is also a hyperradial potential representing the second hyperradial eigenvalue of the adiabatic Hamiltonian in Eq. \eqref{eq:adexpression} that exhibits a potential minimum below $E=0$ which has the potential to contain either bound or resonant states.

\begin{figure}[H] 
\hspace{-0.0in}\includegraphics[width=8.6 cm] {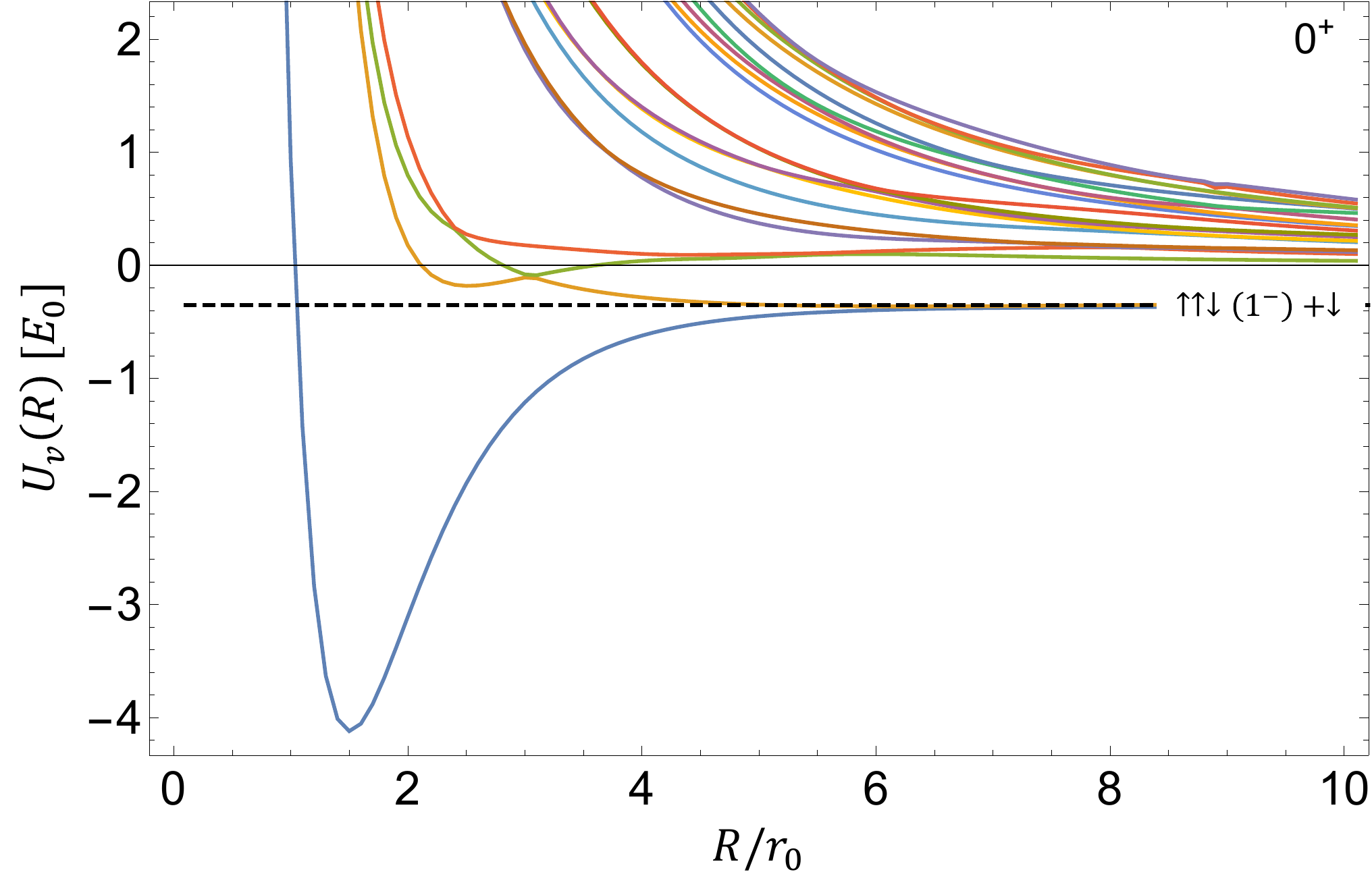}
%\vskip-3.5in 
\caption{\label{fig:4B_uudd_supu} The Born--Oppenheimer potentials for the four--fermion system in the $(\uparrow\uparrow\downarrow\downarrow)$ spin configuration. The two--body interaction between $(\uparrow\uparrow)$ spins is set to the first $p$--wave unitary limit, while the $(\uparrow\downarrow)$ is set to the first $s$--wave unitary limit. For this set of two--body interactions, there is one two--body breakup threshold below the four--body continuum threshold, labeled by the fragmentation of the spins and the angular momentum of the trimer.}
\end{figure}

Next, the hyperradial potentials are mapped out for the $(\uparrow\uparrow\downarrow\downarrow)$ and $(\uparrow\uparrow\uparrow\downarrow)$ spin configurations for total orbital angular momentum $L^{\pi}=0^+$. The lowest few hyperradial potentials for the $(\uparrow\uparrow\downarrow\downarrow)$ configuration are shown in Figure \ref{fig:4B_uudd_supu}. In this spin configuration and symmetry, the lowest two hyperradial potentials converge to a two--body breakup threshold below zero energy that represents the two--body breakup into a trimer plus a free particle in the spin configurations $\uparrow\uparrow\downarrow(1^-)+\downarrow$ and $\downarrow\downarrow\uparrow(1^-)+\uparrow$, where the trimer state is labeled by its total orbital angular momentum, which is $1^{-}$. Through angular momentum coupling, the relative angular momentum between the trimer and free particle in the $0^{+}$ symmetry is $l_{\mathrm{rel}}=1$. In this spin configuration, the lowest hyperradial potential supports one tetramer bound state in the $0^{+}$ symmetry.

\begin{figure}[H] 
\hspace{-0.0in}\includegraphics[width=8.6 cm] {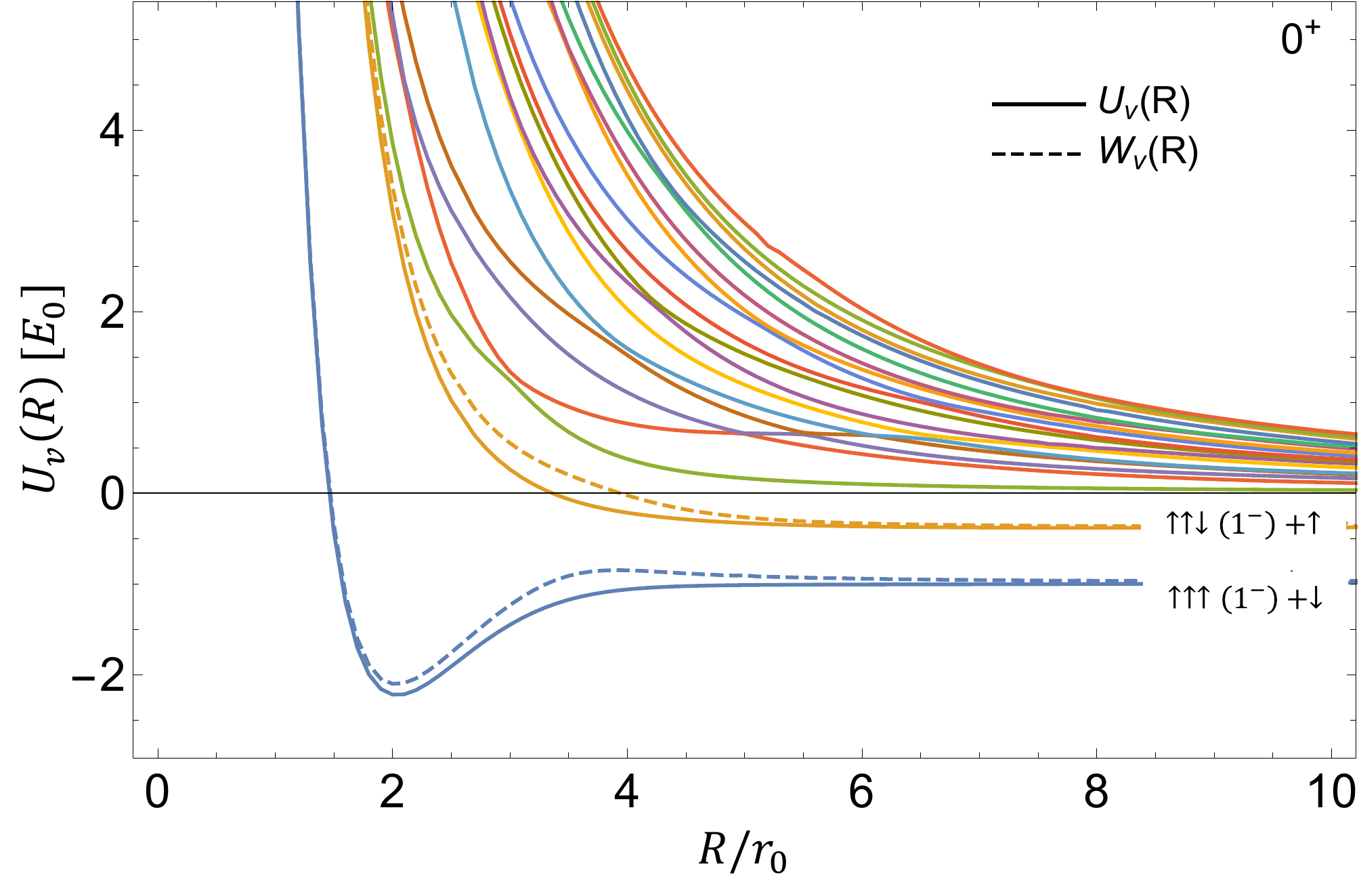}
%\vskip-3.5in 
\caption{\label{fig:4B_uuud_supu} The hyperradial potentials for the four--fermion system in the $(\uparrow\uparrow\uparrow\downarrow)$ spin configuration. The two--body interaction between $(\uparrow\uparrow)$ spins is set to the first $p$--wave unitary limit, while the $(\uparrow\downarrow)$ is set to the first $s$--wave unitary limit. For this set of two--body interactions, there are two two--body breakup thresholds below the four--body continuum, labeled by the fragmentation of the spins and the angular momentum of the trimer. The solid curves are the Born--Oppenheimer potentials and the dashed curves are the effective hyperradial potentials that includes the diagonal non--adiabatic second--derivative coupling.}
\end{figure}

For the other spin configuration, $(\uparrow\uparrow\uparrow\downarrow)$, the hyperradial potentials are shown in Figure \ref{fig:4B_uuud_supu}. In this spin configuration and symmetry, the lowest two hyperradial potentials converge to two--body breakup thresholds below zero energy that represents the two--body breakup into a trimer plus a free particle in the spin configurations $\uparrow\uparrow\downarrow(1^-)+\uparrow$ and $\uparrow\uparrow\uparrow(1^-)+\downarrow$, where again the trimers are in the $1^{-}$ symmetry. In this spin configuration, the hyperradial potentials do not support a tetramer state in this symmetry. In summary, for total orbital angular momentum $0^{+}$, the spin configurations that support tetramer bound states are the $(\uparrow\uparrow\downarrow\downarrow)$ and $(\uparrow\uparrow\uparrow\uparrow)$ configurations where the $(\uparrow\downarrow)$ interactions are tuned to the $s$--wave unitary limit and the $(\uparrow\uparrow)$ and $(\downarrow\downarrow)$ interactions are tuned to the $p$--wave unitary limit.

\section{Conclusions\label{sec:conclusions}}
In this work, the three-- and four--body fermionic systems are studied for various equal--mass cases, interacting near their $s$--wave and/or $p$--wave unitary regimes. Specifically, this work focuses on the regime where there are no two--body bound states (i.e. near the first $s$--wave pole in $a_s$ and the first $p$--wave pole in $V_p$). The symmetries treated in this work are $L^{\pi}=0^{+}$, $1^{-}$, and $2^{+}$ in the spin configurations $(\uparrow\uparrow\downarrow)$ for the three--body system and $(\uparrow\uparrow\downarrow\downarrow)$, $(\uparrow\uparrow\uparrow\downarrow)$, and $(\uparrow\uparrow\uparrow\uparrow)$ for the four--body system. In the first part of this work, the systems considered reside near the $s$--wave unitary limit; our analysis treats universal properties of the hyperradial potentials and the resultant effects on the low--energy elastic phase shifts for continuum scattering. At infinite scattering length, there is a modification to the angular momentum barrier in some of the continuum hyperradial potentials. This modification is followed by a universal long--range next--order term in the hyperradial potentials that falls off as $1/R^3$ and is proportional to the scattering length and is independent of the short-range details of the two--body interaction, and is thus universal. This universal constant is characterised for different symmetries of several three-- and four--body systems in the lowest hyperradial continuum channel. As a result of this long--range universal behavior in the potentials, the low energy phase shift is proportional to the wave number $k$, i.e. to $\sqrt{E}$ in the energy. This leads to a $1/\sqrt{E}$ in the energy--derivative of the phase shift, which is proportional to the density of states in the continuum.

The second part of this article investigates the interaction of equal--mass fermions at the $p$--wave unitary limit for four particles in the symmetry $L^{\pi}=0^{+}$ and its implications as it relates to the properties of the three--body system at $s$--wave and $p$--wave unitary limits where the $s$--wave and $p$--wave dimer energy is at $E=0$. For the spin--polarized case, where the spins are in the $(\uparrow\uparrow\uparrow\uparrow)$ configuration and the $p$--wave interactions are tuned from non--interacting to the unitary regime, the spectrum for the symmetry $L^{\pi}=0^+$ shows one four--body tetramer state supported in the lowest hyperradial potential. From the hyperradial potentials, the potential minimum that supports the tetramer state is located at a distance comparable to the range of the interaction, which is the reason that the ratio of the tetramer energy to the corresponding trimer energy $E_4/E_3$ is sensitive to the form of the short--range behavior of the two--body interaction; this is analogous to what was previously documented for the tetramer states in the lowest hyperradial potential for the four--boson case.

A key quantity for four--body recombination is the ratio of effective $p$--wave scattering lengths associated with occurrence of a zero--energy trimer and tetramer resonance. Through an analysis of Tjon plots for different two--body interactions, this ratio is found to be universal with a value of $a_{p,4}^{0^{+}}/a_{p,3}^{1^{-}}\sim0.88$. This universal property provides a probe for where the loss rate in the four--body recombination process $A+A+A+A\leftrightarrow A_{3}+A$ is enhanced due to the formation of a bound tetramer for the $0^{+}$ symmetry.

\section{Acknowledgements}
We would like to thank both Doerte Blume and Yu-Hsin Chen for fruitful discussions. We would further like to extend thanks to Yu-Hsin Chen for providing some three--body calculations to check with our numerical results. This work is supported
in part by the National Science Foundation, Grant
No. PHY-1912350.

\bibliography{bibliography}
\end{document}